\documentclass[preprint,showpacs,preprintnumbers,amsmath,amssymb,nofootinbib]{revtex4}

\usepackage{etex}
\usepackage{graphicx}% Include figure files
\usepackage{dcolumn}% Align table columns on decimal point
\usepackage{bm}% bold math
\usepackage{amssymb,amsthm,amscd,amsbsy,array}\usepackage{amsmath,dsfont}

\usepackage{graphics,xcolor}

\usepackage{color}
\newcommand{\red}{\textcolor{red}}
\newcommand{\blue}{\textcolor{blue}}

\newcommand{\PT}{{P\"oschl - Teller\;}}

\newcommand{\GW}{{gravitational wave\;}}
\newcommand{\DM}{{Displacement Memory\;}}

\newcommand{\half}{{\scriptstyle{\frac{1}{2}}}}
\def\2{{\half}}
\newcommand{\const}{\mathop{\rm const}\nolimits}
\def\p{{\partial}}
\def\bk{{\bf k}}

\def\bp{{\bm{p}}}

\def\bp{{\bm{p}}}

\def\bX{{\bm{X}}}

\def\beqa{\begin{eqnarray}}
\def\eeqa{\end{eqnarray}}

\def\barray{\left(\begin{array}}
\def\earray{\end{array}\right)}
\def\barraynb{\begin{array}}
\def\earraynb{\end{array}}
\def\smallover#1/#2{\hbox{$\textstyle\frac{#1}{#2}$}} %

\newcommand{\fm}{\mathfrak{m}}

\newcommand{\cA}{{\mathcal{A}}}
\newcommand{\cL}{\mathcal{L}}

\def\smallcirc{{\raise 0.5pt \hbox{$\scriptstyle\circ$}}}
\def\aand{{\quad\text{\small and}\quad}}

\def\where{{\quad\text{\small where}\quad}}

\def\ie{{\;\text{\small i.e.}\;}}
\def\ie,{{\;\text{\small i.e.,}\;}}

\def\GW{{gravitational wave\,}}

\def\VM{{Velocity Effect\,}}
\def\DM{{Displacement Effect\,}}
\def\benu{\begin{enumerate}}
\def\eenu{\end{enumerate}}
\def\bitem{\begin{itemize}}
\def\eitem{\end{itemize}}

\def\besub{\begin{subequations}}
\def\esub{\end{subequations}}
\def\?{{\,\gb{\fbox{\texttt{??}}\;}}\,}

\def\StL{{Sturm-Liouville\,}}
\def\cI{{${\cal I}$}}
\def\cI{{{\cal I}}}

\usepackage[colorlinks=true, pdfstartview=FitV, linkcolor=blue, citecolor=blue, urlcolor=blue]{hyperref} % hyperref

\newcommand{\magenta}{\textcolor{magenta}}
\newcommand{\green}{\textcolor{green}}
\newcommand{\orange}{\textcolor{orange}}

\newcommand{\gb}{\quad\colorbox{green}}

\newcommand{\dgreen}{\textcolor[rgb]{0,0.5,0}}

\newenvironment{redtext}{\color{red}}
{\ignorespacesafterend}
\newenvironment{bluetext}{\color{blue}}{\ignorespacesafterend}
\newenvironment{greentext}{\color{green}}{\ignorespacesafterend}
\newenvironment{magentatext}{\color{magenta}}{\ignorespacesafterend}
\newenvironment{cyantext}{\color{cyan}}{\ignorespacesafterend}
\newenvironment{orangetext}{\color{orange}}
{\ignorespacesafterend}

\newcommand{\bmagenta}{\begin{magentatext}}
\newcommand{\emagenta}{\end{magentatext}}
\newcommand{\bcyan}{\begin{cyantext}}
\newcommand{\ecyan}{\end{cyantext}}
\newcommand{\bblue}{\begin{bluetext}}
\newcommand{\eblue}{\end{bluetext}}
\newcommand{\bred}{\begin{redtext}}
\newcommand{\ered}{\end{redtext}}

\newcommand{\bgreen}{\begin{greentext}}
\newcommand{\egreen}{\end{greentext}}
\newcommand{\borange}{\begin{orangetext}}
\newcommand{\eorange}{\end{orangetext}}

\numberwithin{equation}{section}

\let\ssection=\section
\renewcommand{\section}{\setcounter{equation}{0}\ssection}
\newcommand{\beq}{\begin{equation}}
\newcommand{\eeq}{\end{equation}}
\newcommand{\bec}{\begin{center}}
\newcommand{\ec}{\end{center}}

\newcommand{\medbox}[1]{\fbox{%
\rule[-10pt]{0pt}{25pt}$\;\;\displaystyle{#1}\;\;$}%
}

\usepackage{amsthm}

{Corollary}[section]

\begin{document}

\preprint{arXiv:2609.39694-v3}

\title{Gravitational Memory:  
\\
an
\\
 approximate description\footnote{\small Presented by Q-L Zhao and  by P. Horvathy at the  
 14th Bolyai-Gauss-Lobachevsky (BGL-2026) Conference. %Sept 10 2026
 }}

\author{
Q.-L. Zhao $^{1}$\footnote{mailto:zhaoqliang@ucas.ac.cn},
P.-M. Zhang$^{2,3}$\footnote{Corresponding author mailto:zhangpm5@mail.sysu.edu.cn},
M. Elbistan$^{3}$\footnote{mailto:mahmut.elbistan@bilgi.edu.tr},
J. Balog$^{3,4}$\footnote{mailto:balog.janos@wigner.hu
}
and
P.~A. Horvathy$^{3,5}$\footnote{mailto:horvathy@univ-tours.fr}
}

\affiliation{
${}^{1}$ School of Fundamental Physics and Mathematical Sciences,
		Hangzhou Institute for Advanced Study, UCAS, Hangzhou 310024, China	
\\
$^2$ School of Physics and Astronomy, Sun Yat-sen University, Zhuhai 519082, (China)
\\
${}^3$ Department of Energy Systems Engineering, Istanbul Bilgi University, 34060, Eyupsultan, Istanbul, (Turkey)
\\
${}^{4}$ Holographic QFT Group, Institute for Particle and Nuclear Physics,
HUN-REN Wigner Research Centre for Physics
H-1525 Budapest 114, P.O.B. 49, (Hungary),
\\
${}^{5}$ Institut Denis-Poisson CNRS/UMR 7013 - Universit\'e de Tours - Universit\'e d'Orl\'eans Parc de Grammont, 37200; Tours, (France).}
%%%%%%%

%
\begin{abstract}
Plane gravitational waves play a  role both in relativistic and in non-relativisitic physics, highlighted by two different mass parameters.
The large-distance behaviour of a sandwich gravitational wave can be approximated by a continuous but not necessarily smooth profile, providing us with a simplified description of particle motion. Our model is consistent with the Carroll symmetry. Our strategy is illustrated by the P\"oschl-Teller profile. 
The limitations of the sandwich approximation are discussed. %
\end{abstract}

%\pacs{}

\maketitle

\tableofcontents

%%%%%%%%%%%%%%%%%%%%%%%%%%%%%%%%%%%%%%%%%%%%%%%%%%%%%%%%%%%%%%%%%%%%%%%%%%%
\section{Introduction}\label{Intro}
%%%%%%%%%%%%%%%%%%%%%%%%%%%%%%%%%%%%%%%%%%%%%%%%%%%%%%%%%%%%%%%%%%%%%%%%%%%

\begin{figure}[h]
\includegraphics[scale=.25]
{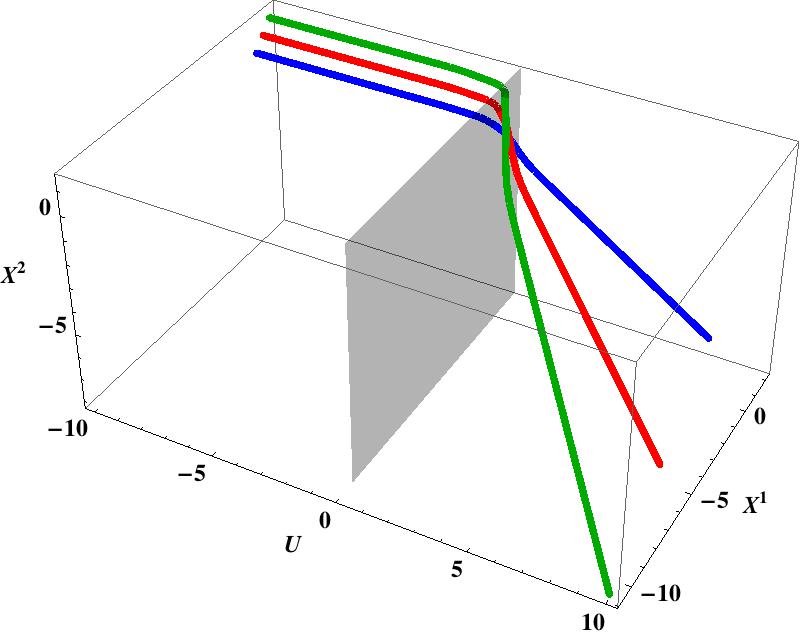}
\vskip-5mm\caption{\textit{\small Particles initially at rest hit by a burst of gravitational waves fly apart with constant velocity following diverging trajectories, consistently with the Velocity Memory Effect (VM). 
}
\label{3dtrajectory}
}
\end{figure}

An early proposal to detect gravitational waves suggested to observe the displacement of particles 
 by a passing wave, called the \emph{Memory Effect} (ME)  \cite{BraTho}. 
Initial study \cite{Ehlers} argued in favour of the \emph{\VM} (VM)~: particles at rest before being hit by the wave would fly off with \emph{non-zero constant velocity}, as shown in FIG.\ref{3dtrajectory}. 

%%%%%%%%%%%%%%%%%% FIG 3dtrajectory
%%%%%%%%%%%%%

Zel'dovich and Polnarev \cite{ZelPol} argued instead  that the particles would merely be displaced, consistently with the \emph{Displacement Effect} (DM).

%%%%%%%%%%%%% FIG Scarf3D
\begin{figure}[h]
\includegraphics[scale=.27]{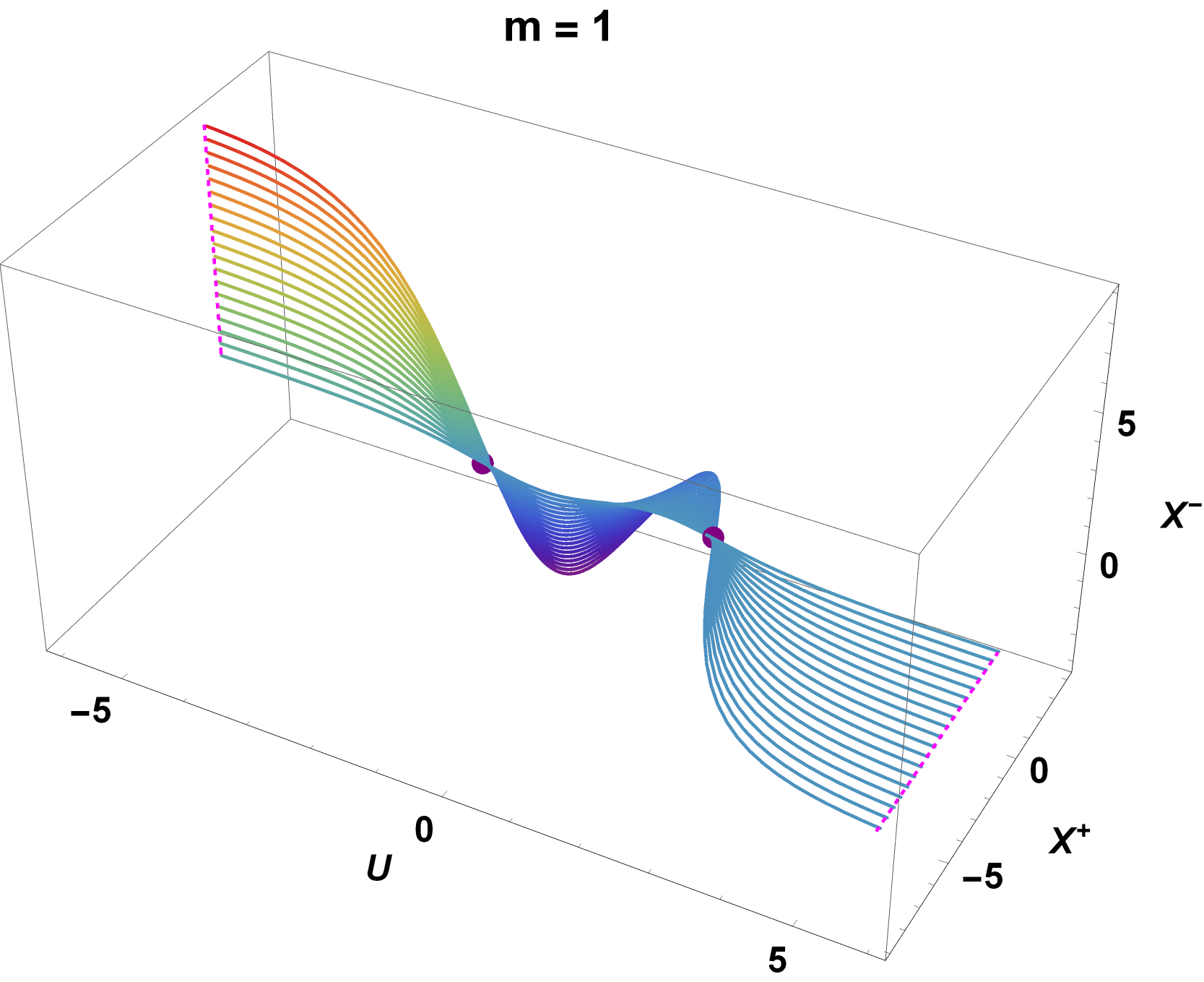}
\quad
\includegraphics[scale=.27]{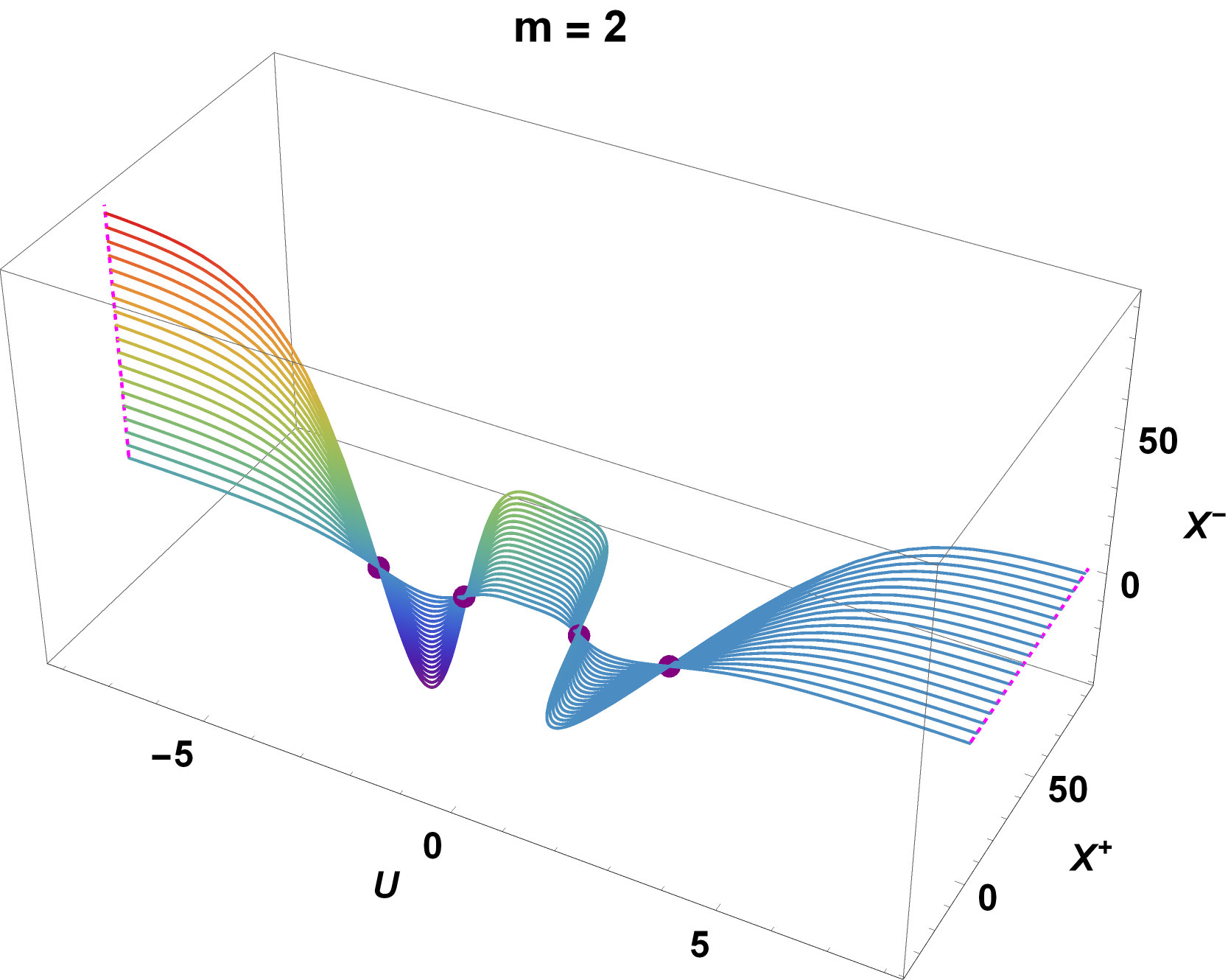}

\vskip-5mm
\caption{\textit{\small
After passing through the Wavezone, the 
DM trajectories with wave number ${\bf m}$ are rotated  by 
 \red{${\bf (-1)^m\frac{\pi}{2}}$}\,. 
 }
\label{Scarf-m1m2-DM}
}
\end{figure}
%%%%
Their statement could be confirmed (theoretically), for 
 various ``sandwich'' wave profiles~\footnote{While the intuitive meaning of a ``sandwich wave'' is clear, the concept requires precise growth conditions \cite{Klisch,ZhaoCao}, though, as discussed in Sec.\ref{LimitMemSec}.}, including a Gaussian  or a \PT potential \cite{LongMemory,EZHRev}, their derivatives \cite{PTeller, Chakra, DM-1,DM-2} or for the Scarf  \cite{Scarf,Sila-PLB}, and even for a simple square profile \cite{Kar3,Benin} --- \emph{provided the wave amplitude takes some ``magic'' value $k_{crit}$} \cite{DM-1,DM-2,Jibril,Benin}.
 
The models mentioned above require knowledge of special function theory.  A curious observation says, though, that \emph{different} wave profiles can yield rather similar behaviour, as exemplified by the Gaussian,
 \begin{equation}
\cA^{Gauss}(U) = \frac{k}{\sqrt{\pi}}e^{-U^{2}}\,
\label{GaussProf}
\eeq
(which has only numerical solutions) and the
{\PT} profile \cite{PTeller},
\beq
\cA^{PT}(U) = \dfrac{m(m+1)}{\cosh^2 U}\,,
\label{PTellerPot}
\eeq
for which the geodesic eqns can be solved analytically.
However, a surprisingly efficient  \emph{approximate} treatment which bypasses the technical difficulties was proposed recently \cite{Approxi}, illustrated here by the \PT profile.
\goodbreak

%%%%%%%%%%%%%%%%%%%%%%%%%%%%%%%%%%%%%%%%%%
\section{Memory effect in a plane \GW} \label{MemoGWSec}
%%%%%%%%%%%%%%%%%%%%%%%%%%%%%%%%%%%%%%%%%%

We consider a linearly polarised vacuum wave with Brinkmann  metric,
\beq
g_{\mu\nu} dX^{\mu} dX^{\nu} = d\bX^2+2dUdV + \cA(U)\Big((X^{+})^2-(X^{-})^2\Big)
\,dU^2\,,
\label{Bplanewave}
\eeq
where the
 $\bX = (X^+,X^-)$ are coordinates of the transverse plane carrying a flat Euclidean metric $d\bX^2=\delta_{ij}\,dX^idX^j$; $U$ and $V$ are light-cone coordinates.
 $\cA(U)$ is  the profile of the wave.
 The relative minus follows from the vacuum Einstein equations. 
 The eqns of motion of a scalar particle are, \cite{Carrollvs,Carroll4GW}. 
%  
%%%%%%%%%%%%%%%%%%%%%%%%%%
\begin{subequations}
\begin{align}
&{\dfrac {d^2\!X^+}{dU^2} - \cA(U) X^+ = 0\,,} \label{geoX1}
\\[6pt]
&{\dfrac {d^2\! X^-}{dU^2} + \cA(U)  X^- = 0\,,}
\label{geoX2}
\\[8pt]
&
\dfrac {d^2\!V}{dU^2}
+\frac{1}{2}\dfrac{d\cA}{dU}\Big((X^+)^2-(X^-)^2\Big)
 +
2\cA(U)\Big(X^+\frac{dX^+}{dU}-X^-\dfrac{dX^-}{dU}\Big)=0\,.
\label{geoVfly}
\end{align}
\label{Bgeoeqn2}
\end{subequations}
%%%%%%%%%%%%%%%%%%
%
Eqn. \eqref{geoVfly}  corresponds to the null lift of $\bX$ and thus follows from \eqref{geoX1} - \eqref{geoX2}. For $\cA(U) >0$ the $X^+$ sector is repulsive and that of $X^-$ is attractive.

The eqns \eqref{Bgeoeqn2} should be supplemented by fixing the value of the conserved quantity called the \emph{Jacobi invariant}
\cite{exactsol},
%%%%%
\beq
\mathfrak{m}^2 = g_{\mu\nu}\dot{X}^{\mu}\dot{X}^{\nu} = \const .
\label{Jacobiinv}
\eeq
%%%%%
viewed here as a constraint~: $\mathfrak{m}^2<0$ for a massive, and $\mathfrak{m}^2=0$ for a lightlike particle. 
($\fm^2 > 0$ would correspond to 
tachyons, which are not considered here).
The trajectory \eqref{geoX1}-\eqref{geoVfly}
is independent of $\fm$ \cite{Eisenhart,DBKP,DGH91}.  

Our theory has actually a second mass parameter. The Killing vector $\p_V$ of \eqref{Bplanewave} generates a conserved quantity, $M$, and the relativistic theory projects, in the  E-D framework \cite{Eisenhart,DBKP,DGH91}, to a non-relativistic one in one dimension lower. Null geodesics project in particular to non-relativistic motions %
 with $M$ playing the role of non-relativistic mass. 
\beq
{\cL}_{NR} = 
    \frac{M}{2}\left(
    \big(\frac{d\bX}{dU}\big)^2
    +\cA(U)\Big((X^+)^2-(X^-)^2\Big)\right)
    \,.
    \label{NRL}
\eeq  
Both the relativistic and the non-relativistic masses will play important roles. For simplicity, henceforth we only consider the attractive $D=1$ sector, $X \equiv X^{-}$.

For a  particle at rest in transverse space  before being hit by the wave the initial conditions are,
\beq
X(-\infty)= X_{0}=\const
\qquad
\frac{dX}{dU}(-\infty) = 0\,.
\label{Xinitcond}
\eeq
The \emph{\DM\!} (DM) arises when we have, in addition,
\beq
\frac{dX}{dU}(+\infty) = 0\,
\label{DMboundcond}
\eeq
which implies $X(+\infty)= \const$.~\footnote{
A sufficient condition is that profiles decay as $1/U^p, \, p>3$ \cite{Klisch, ZhaoCao}, see sect.\ref{LimitMemSec}.}
The system \eqref{Xinitcond}--\eqref{DMboundcond} is clearly overdetermined and solutions do exist only for particular wave amplitudes $k=k_{crit}$ identified either numerically, or by using  advanced knowledge of special function theory, as confluent Heun functions \cite{DM-2}, or by sophisticated methods as the Nikiforov-Uvarov algorithm \cite{Sila-PLB}.
%%%%%%%%

%%%%%%%%%%%%%%%%%%%%%%%%%%%%%%%%%%%%%%%%%%
\section{Approximate profile}\label{ApproxToy}
%%%%%%%%%%%%%%%%%%%%%%%%%%%%%%%%%%%%%%%%%%

The curious independence of the concrete form of the profile \cite{LongMemory,EZHRev,PTeller, Chakra, DM-1,DM-2,Sila-PLB} lead us to speculate \cite{Approxi}.
Here we spell out our point for  the \PT profile \eqref{PTellerPot} 
with wave amplitude $k^2=m(m+1)$  \cite{Chakra,DM-1,DM-2}.
The \StL eqn \eqref{geoX2} 
satisfies the DM condition \eqref{DMboundcond}
 when $m$ is a positive integer.
The trajectories are then composed of $m$ half-waves \cite{DM-1,DM-2}. 

An approximate ``toy'' model \cite{Approxi}  is obtained as follows. For large $U$, $\cosh^{-2}(U)\approx 4e^{-2|U|}$,
and we approximate  \eqref{PTellerPot}, for $m=1$, by,
%%%%%%
\beq
\cA^{approx}(U) = k^2e^{-2|U|}\,,
\quad
k \approx 2.4 \dots
\label{extoy}
\eeq
%%%%%% 
 For large $|U|$, the profile is close to that of PT in \eqref{PTellerPot}, as seen in FIG.\ref{exp-PT}. The approximation breaks manifestly down in the neighbourhood of the origin.

%%%%%%%%%%%%%%%%% FIG exp-PT
\begin{figure}[h]\quad
\includegraphics[scale=.8]{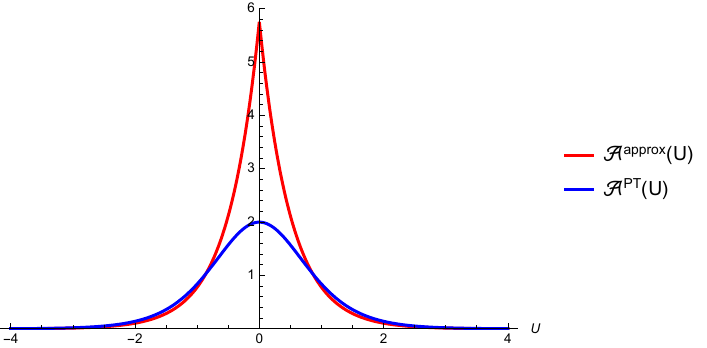}
\vskip-3mm\caption{
 \textit{\small For large $|U|$, the \red{\bf approximate profile} \eqref{extoy} with amplitude $k\approx 2.4$ is close to that of \blue{\PT\!}  $\cA^{PT}$ with $m=1$, in \eqref{PTellerPot}.
}
\label{exp-PT} }
\end{figure}
%%%%%%%%%%%%%%%%%

Our approximate model has exact geodesics, given by a combination of Bessel functions \cite{ZZH22}.
The DM boundary conditions $X^{\prime}(\pm\infty)=0$  
leave us with
\beq
X(U) = \left\{\barraynb{llcll}
X_{+}(U)&=& \alpha_1J_0\big(ke^{-U}\big)  &\text{in}\;& \cI_{+}= \big\{U > 0\big\}
\\[4pt]
X_{-}(U)&=&\alpha_2\,J_0\big(ke^{U}\big)  &\text{in}\;& \cI_{-} = \big\{U <0\big\}
\earraynb\right.
\,,
\label{regBessel}
\eeq
where $\alpha_1$ and $\alpha_2=X_0$ are arbitrary constants.
Admissible trajectories are obtained when the left and right solutions match smoothly at $U=0$ which is manifestly not so for randomly chosen amplitude, as shown in FIG. \ref{discont}. 
%
%%%%%%%%%%%%%%%%%%% FIG discont
\begin{figure}[h]
\includegraphics[scale=.6]{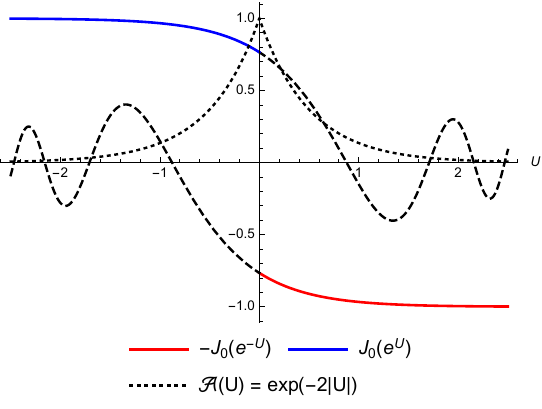}
\qquad\;
\includegraphics[scale=.6]{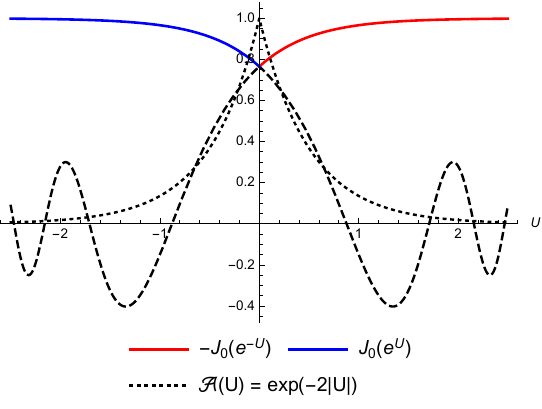}

\vskip-4mm\caption{\textit{\small For amplitude  $k\neq k_{crit}$
 the positive and negative $U$-branches do not match smoothly~:
for  antisymmetric fitting  $X_{-}(0)\neq X_{+}(0)$  and  for  symmetric fitting $X_{-}'(0)\neq X_{+}'(0)$.
}
\label{discont} }
\end{figure}
%%%%%%%%%%%%%%%%%%%

Increasing the amplitude pulls apart the left and right branches as shown in FIG.\# 5 of \cite{Approxi};  fitting them provides us with two possibilities~:
\benu
\item
When  
$ J_0(k)=0$ 
we get an \emph{odd} wave number, $m = 2\ell+1$.
The two branches are joined at the origin and the slopes must be equal, 
\beq
\label{oddcondition}
  X_-(0)=X_-(0)=0\,,
\qquad
X_{-}'(0)=X_{+}'(0)\,,
\eeq
which then requires
to  glue the $U < 0$ and $U > 0$ branches antisymmetrically
%%%%%
\beq
X^{odd}(U) = \left\{\barraynb{llcll}
X_{+}(U)&=& - \alpha\, J_0\big(ke^{-U}\big)  &\;\text{in}\;& \cI_{+}
\\[3pt]
X_{-}(U)&=&\,\alpha \,J_0\big(ke^{U}\big)  &\;\text{in}\;& \cI_{-}
\earraynb\right.
\,.
\label{DModd}
\eeq
%%%%%

%%%%%%%%%%%%%%%
Increasing the amplitude pulls the branches apart symmetrically \cite{Approxi}. Smooth matching is obtained 
when the left and right branches match smoothly at $U=0$, which happens for $k=k_{crit}$. DM is obtained when 
\beq
 \text{either} \;\; X_{-}(0)=X_{+}(0) = 0
\quad\,
\text{or} \;\;\;\; (X_{-})^{\prime}(0)=(X_{+})^{\prime}(0)
 = 0\,.\quad
\eeq
\goodbreak

Despite the lack of smoothness of the profile, 
the approximate trajectories \eqref{DModd} are surprisingly close to those of full  \PT   with
odd half-wave number $m=2\ell+1$, as seen in FIG.\ref{modd}. 
%%%%%%%%%%%%%%%%%%% FIG Toy-odd
\begin{figure}[h]
\includegraphics[scale=.27]{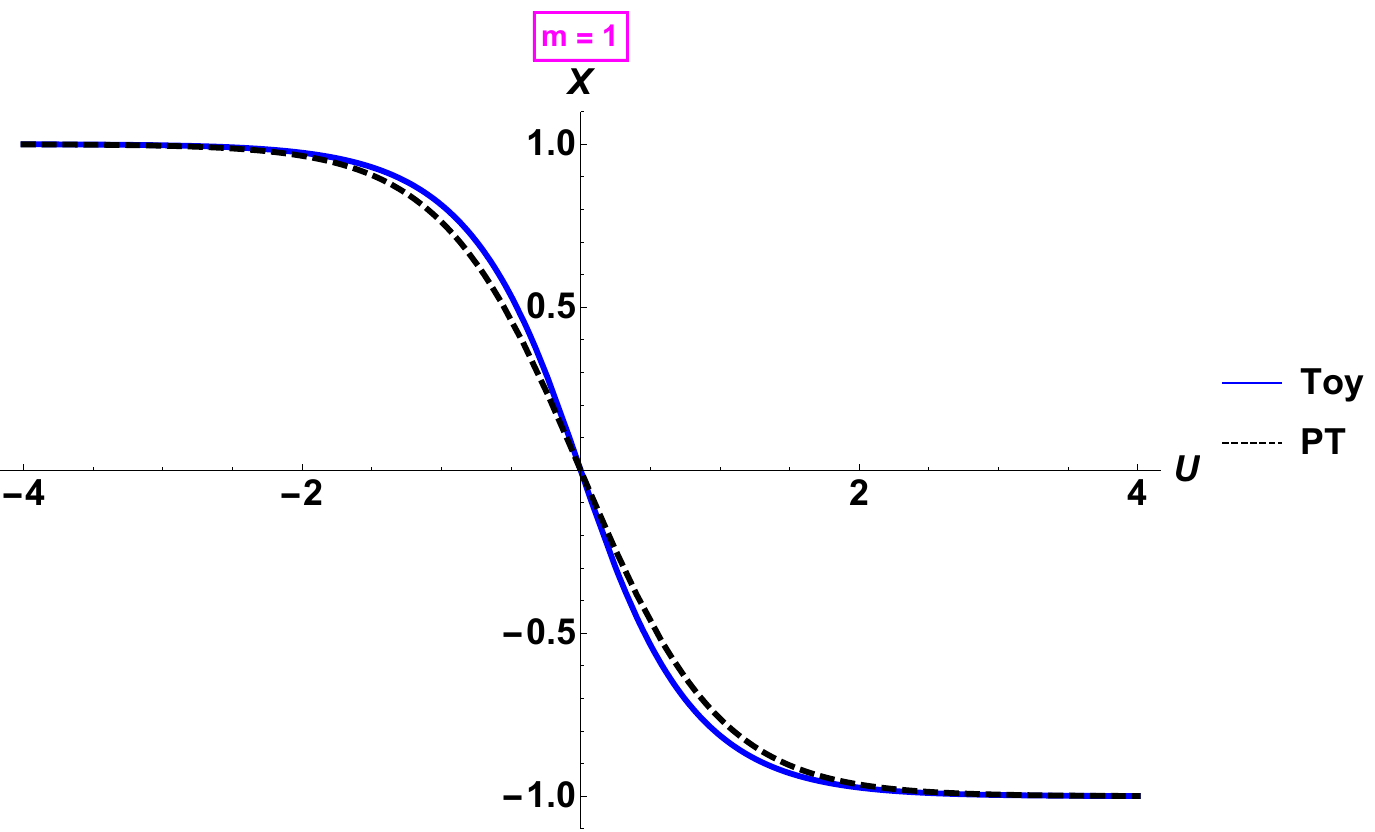}\quad\;
\includegraphics[scale=.27]{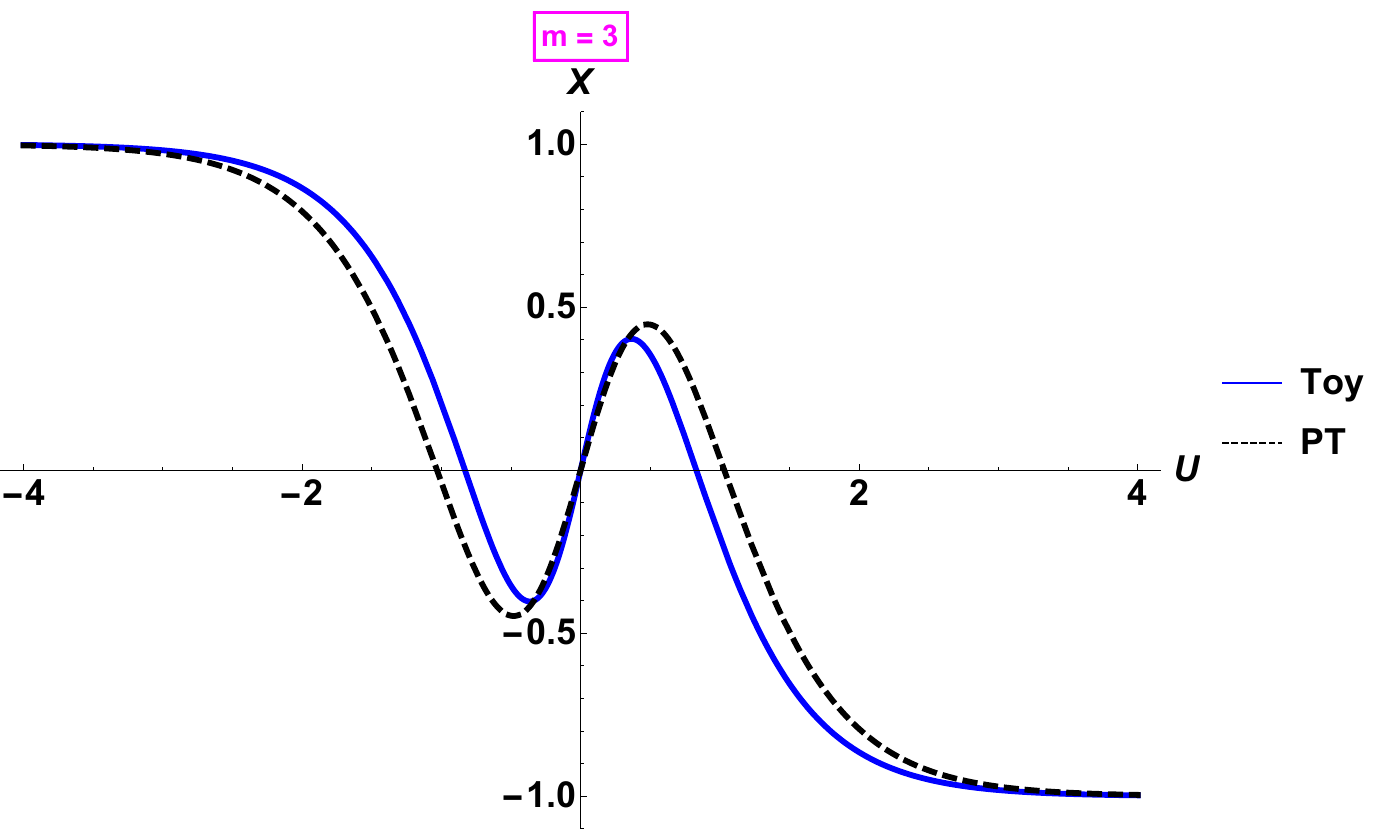}

%\vskip-1mm
%\hskip-13mm (a) \hskip72mm (b)
\vskip-4mm \caption{\textit{\small
The DM  trajectories (dashed) for the {\bf toy model} \eqref{extoy}
 approximate  those of \, \blue{\bf \PT}  with {\bf odd} half-wave number \red{$m=2\ell+1$}.}
\label{modd} }
\end{figure}
%%%%%%%%%%%%%%%%%%%

\item 
Another possibility with $J_0(k)\neq0$ is obtained by requiring that the trajectory be smooth at the junction, where $X_{-}(0)=J_0(k)=X_{+}(0)$,
\beq
X'(0)=0.
\label{Xprime0}
\eeq
 Then the DM trajectory with initial position $X_0$, 
\beq
X^{even}(U) = J_0\big(k_{(2\ell)}e^{-|U|}\big)\,X_0 \,,
\label{eventraj}
\eeq
is a good approximation of  \PT  for \emph{even} half-wave number $m=2\ell$. It
is obtained by gluing together the negative and positive $U$-branches \emph{symmetrically},
$ 
X(-U) =  X(U)\,,
$ 
yielding ``DM trajectories with no displacement'', shown in FIG.\ref{meven}. 
\eenu
%%%%%%%%%%%%%%%%%%% FIG Toy-even
\begin{figure}[h]
\includegraphics[scale=.27]{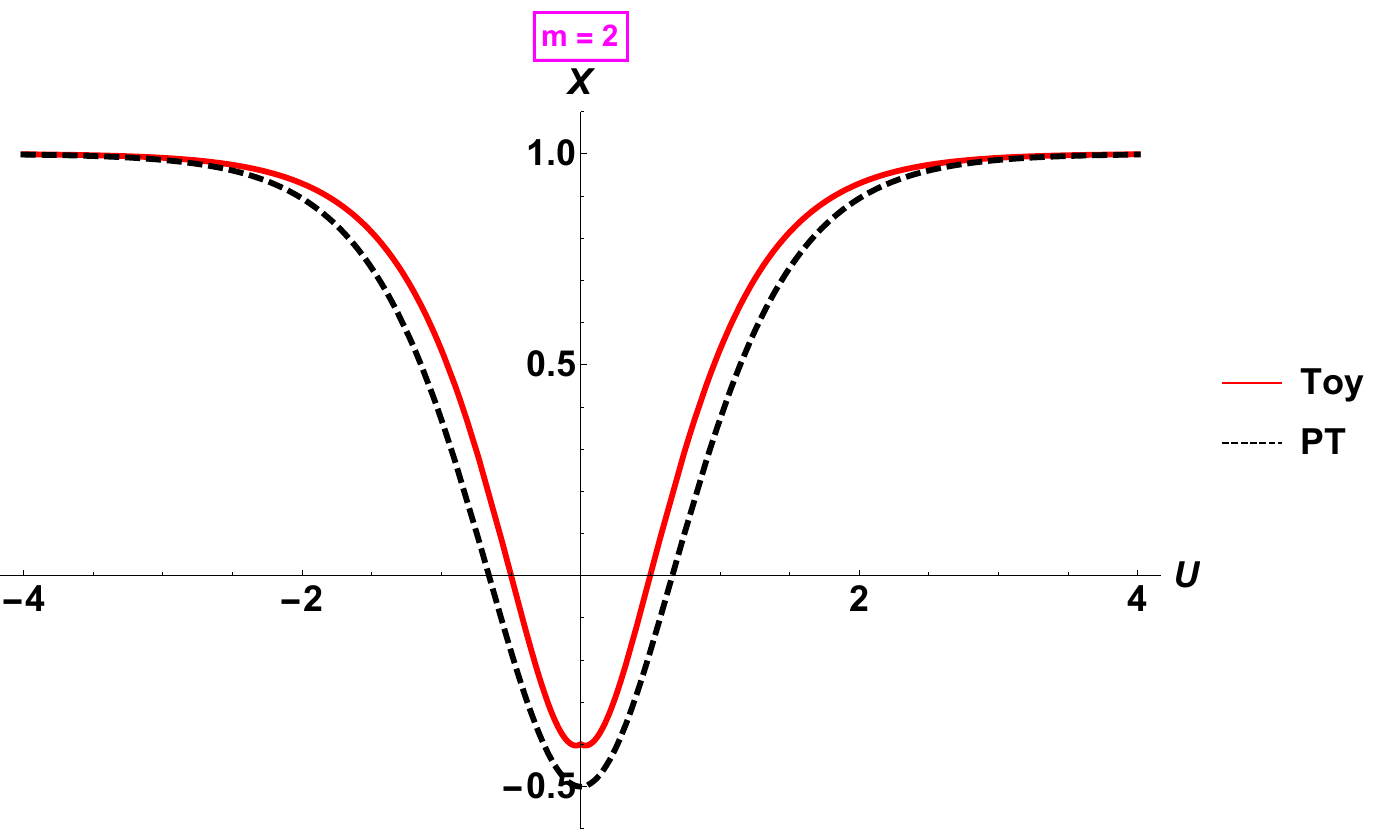}\;\;\;
\includegraphics[scale=.27]{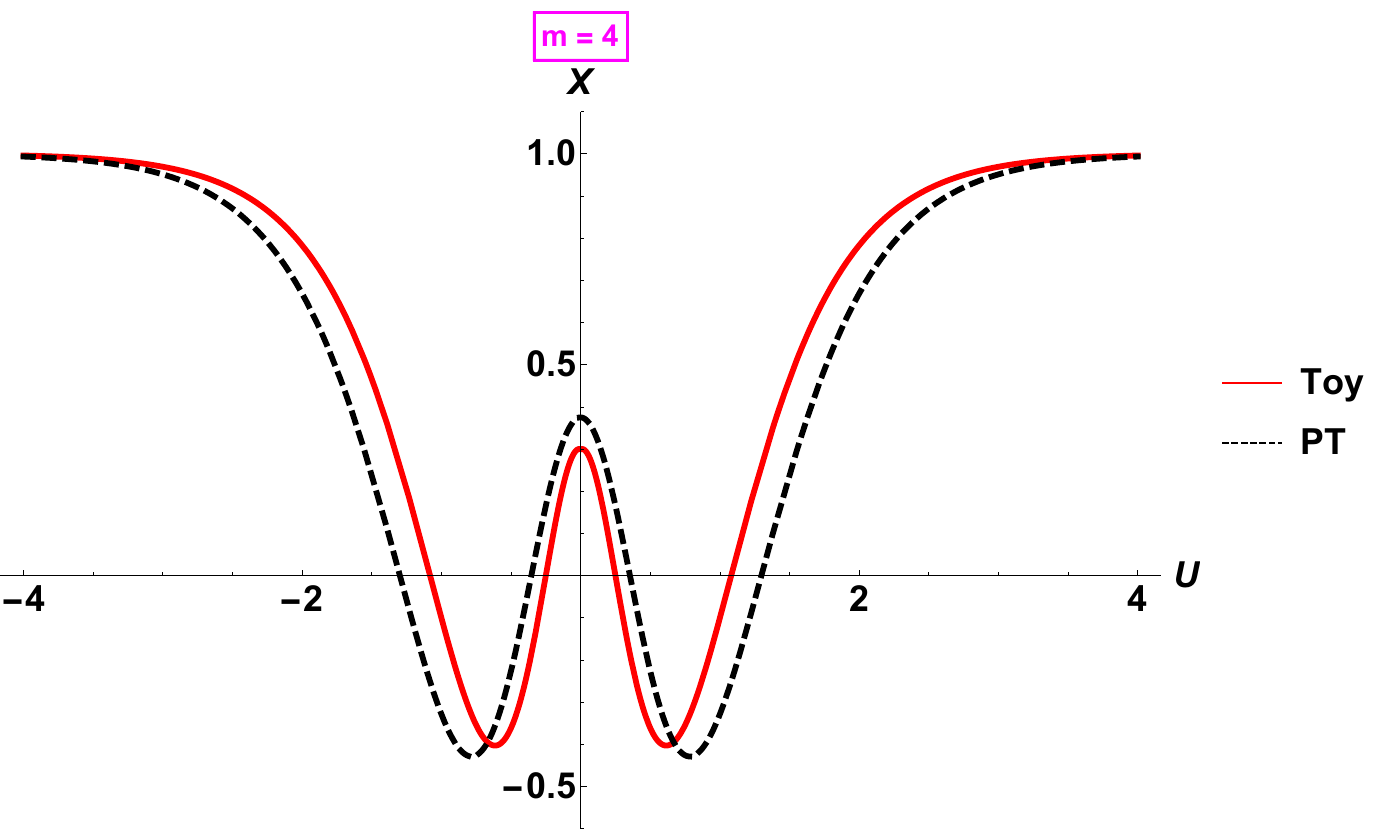}

%\vskip-2mm
%\hskip-11.5mm (a) \hskip70.5mm (b)
\vskip-5mm \caption{\textit{\small
DM toy trajectories (dashed) are obtained  when the 
 Wavezone contains an {\bf even} number, \red{$m=2\ell$}, of symmetrically glued half-waves. They approximate those for \red{\bf \PT\!}.
 }
\label{meven} }
\end{figure}
%%%%%%%%%%%%%%%%%%%

%%%%%%%%%%%%%%%%%%%%%%%%%%%%%%%%%%%%%%
\section{Geodesics}\label{GeodesicSec}
%%%%%%%%%%%%%%%%%%%%%%%%%%%%%%%%%%%%%%

Now we return to the general theory of sect.\ref{MemoGWSec}. 
Our clue is that for DM parameters the total Hamiltonian action of the projected NR model vanishes by the boundary conditions \eqref{Xinitcond}-\eqref{DMboundcond} \cite{DM-1},
\beq
\label{DMaction}
\displaystyle\int_{-\infty}^{\infty}\!\!\cL_{NR}\, dU \; \propto \;
\frac{1}{2}XX^{\prime }\Big|_{-\infty}^{\;\infty}=0\,
\eeq
for all values of the Jacobi invariant $\fm$. However the vertical coordinate, $V$, behaves somewhat differently in the massless and massive cases, as we discusss it now.
%%%%%%%%%%%

%%%%%%%%%%%%%%%%%%%%%%%%%%%%%%%%%%%%%%
\subsection{Massless geodesics}\label{MasslessSec}
%%%%%%%%%%%%%%%%%%%%%%%%%%%%%%%%%%%%%%
In the massless case, the vertical coordinate is obtained by lifting the transversal trajectory $X(U)$ horizontally to Bargmann space,
\beq
{V}(U)=
 V_0 - \frac{1}{M}\cI(U)\,, \qquad \cI(U) =
 \int_{-\infty}^U\!\!{\cL}_{NR}\,du\,.
\label{nullV}
\eeq  
For DM parameters
$ 
\cI(+\infty) =
\int_{-\infty}^{\;\infty}\!\cL_{NR}\,du=0\, 
$ 
 is thus eliminated, leaving us with,
\beq
V_{(out)} = V_0 = V_{(in)}\,\,
\label{nullVV0}
\eeq 
%%%%%%%%%%%%%%%%%%%%%%%%
extending DM to all coordinates, 
as shown  in FIG.\ref{PTXV}.
%%%%%%%%%%%%%%%%%% FIG PT-XV-m1m2
\begin{figure}[h]
\includegraphics[scale=.3]{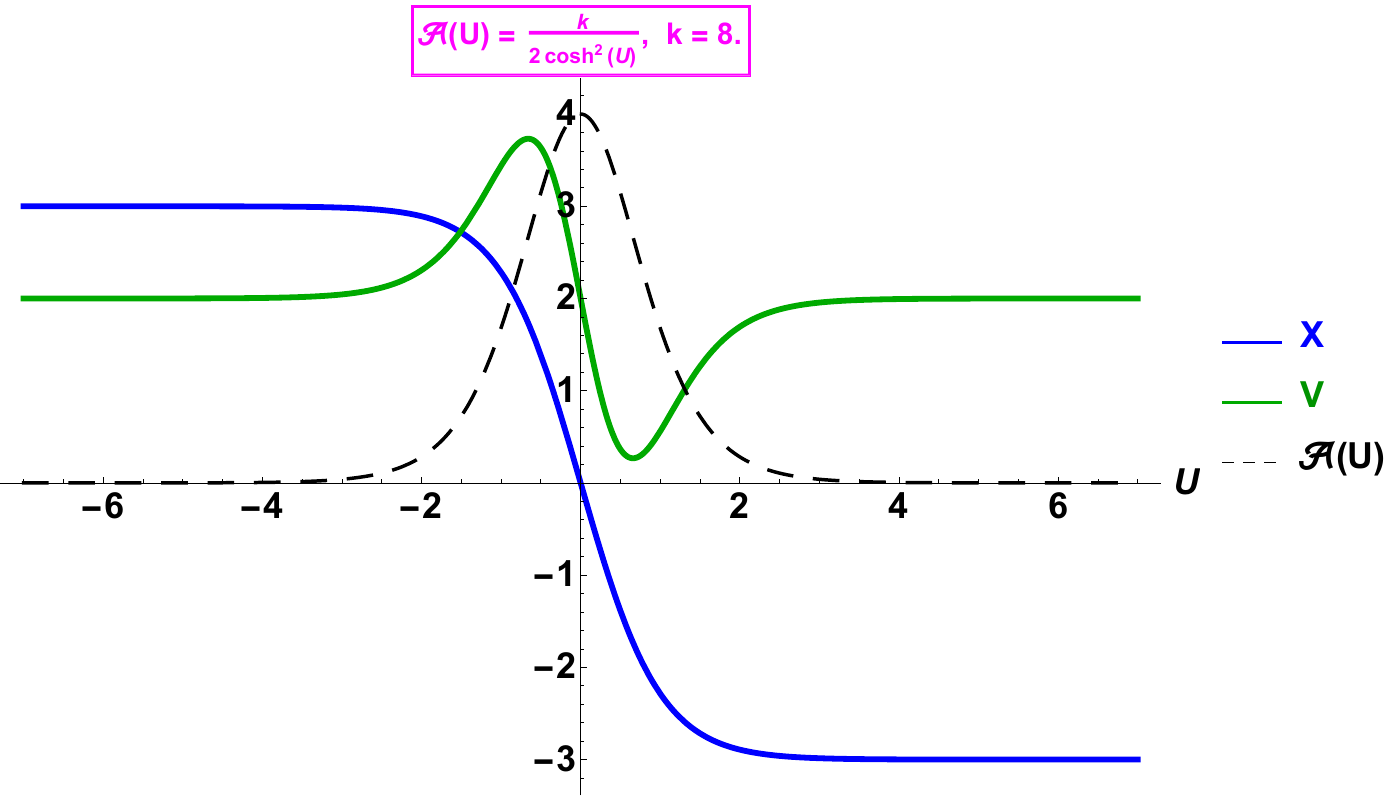}
\hskip4mm
\includegraphics[scale=.3]{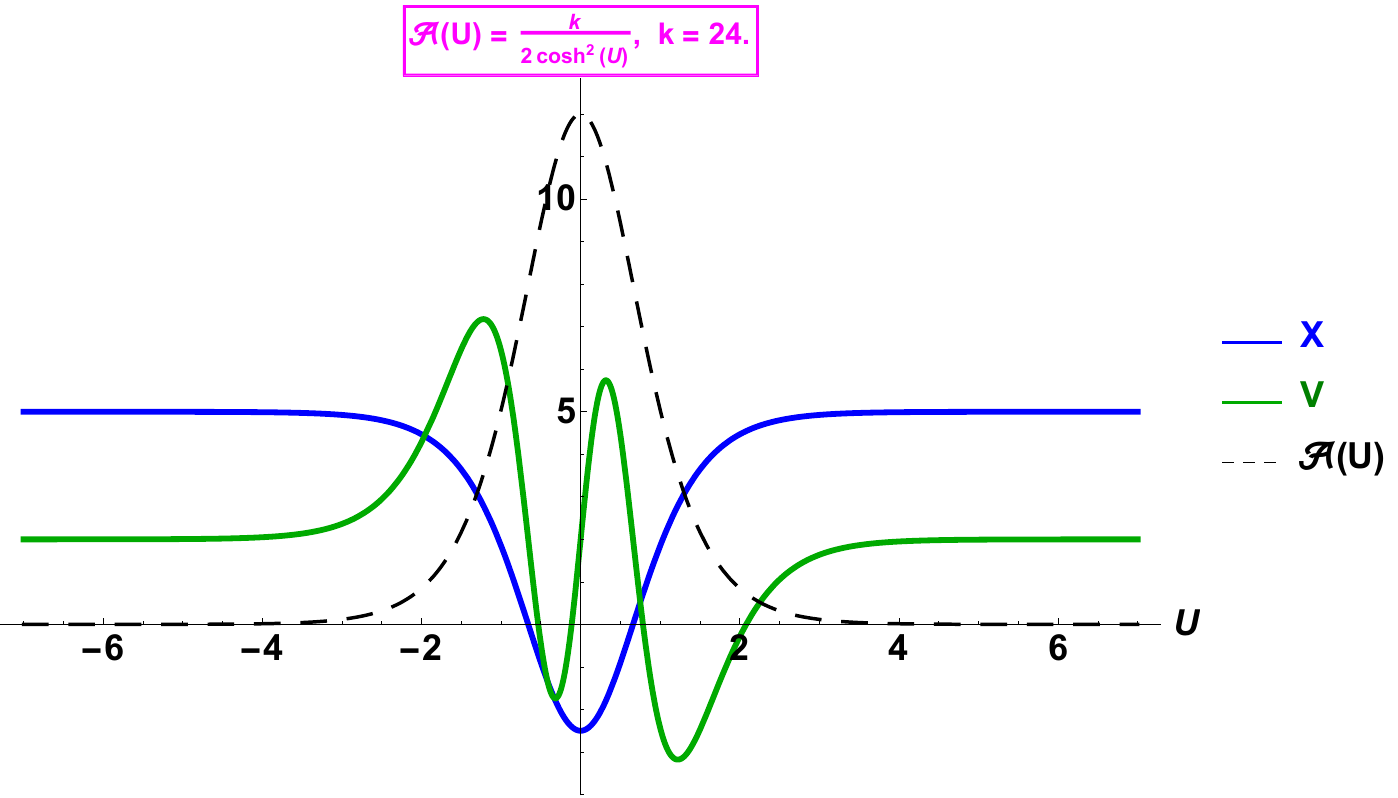}

%\\
%\vskip-2mm\hskip-10mm (a) \hskip 71mm (b)
\vskip-5mm
\caption{\textit{\small For critical amplitude $k=k_{crit}$
both  the \blue{\bf transverse, $X$}, and  the \dgreen{\bf vertical, $V$}, coordinates exhibit DM, as illustrated here for \PT in $D=1$, with ${\bf \magenta{m=1,\,2}}$.}
\label{PTXV}
}
\end{figure}
%%%%%%%%%%%%%%

%%%%%%%%%%%%%%%%%%%%%%%%%%%%%%%%%%%%%%
\subsection{Massive geodesics}\label{MassiveSec}
%%%%%%%%%%%%%%%%%%%%%%%%%%%%%%%%%%%%%%

Now we consider the physically relevant particles with nonzero relativistic mass, $\mathfrak{m}\neq0$ \cite{Andrz20}. Integrating twice the $V$ eqn \eqref{geoVfly} augmented by the constraint $\mathfrak{m}^2 = \const < 0 \,,$
%%%%%
\beq
V^{\fm}(U)=V_{0}
-\frac{1}{M}
\int_{-\infty}^U \cL_{NR}(y) dy\;
 + \frac{1}{2}\frac{\mathfrak{m}^{2}}{M^{2}}\,U \,, 
\label{massiveVBIS}
\eeq
%%%%%%%
yields, by using \eqref{DMaction},
%%%%%%%%%%%%
\beq
V^{\mathfrak{m}}(U)=V^{\fm=0}(U)
 + \;\frac{1}{2}\frac{\mathfrak{m}^{2}}{M^{2}}\,U \,.
\label{massiveV}
\eeq
%%%%%%%
Thus $V$ picks up a linear-in-$U$ term,
 which tilts the $V(U)$ trajectories, FIG.\ref{Vmplot}.
\goodbreak

%%%%%%%%%%%%%%%%%%%%%%%  FIG VM plot
 \begin{figure}[h]
\includegraphics[scale=.38]{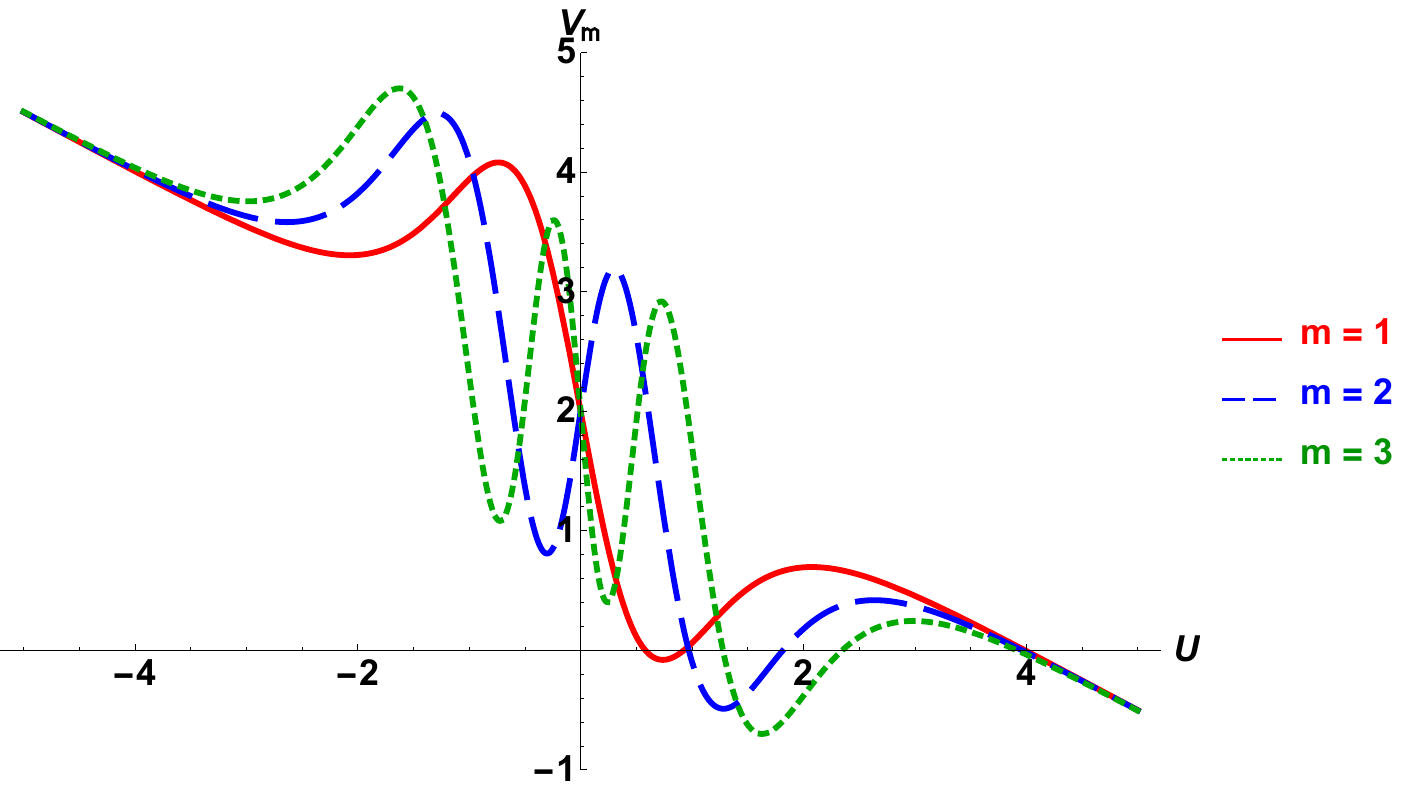}
\vskip-5mm \caption{\textit{\small In the massive relativistic case $\mathfrak{m}\neq0$ the ``vertical'' coordinate $V$  exhibits VM.}
\label{Vmplot} }
%%%%%%%%%%%%%%%%%%%%%%%
\end{figure}
 
Does the mass break DM ? We argue that no, it does \emph{not}. 
Switching indeed to longitudinal coordinates \cite{DM-1}, 
%%%%
\beq
Z = V +\half U\,,\qquad T =  V -\half U
\label{ZVU}
\eeq
%%%%
we find $dZ^2 - dT^2 =  2dU dV$, and
%%%%%%
\beq
Z(U) = V^{\fm=0}(U)+\frac{1}{2}\left(1+\frac{\mathfrak{m}^2}{M^2}\right)\,U\,.
\label{ZUV}
\eeq
%%%%
Then choosing 
\beq 
\mathfrak{m}^2 = - M^2
\eeq
%%%%
the $U$-term in \eqref{ZUV} is eliminated and the 
trajectory is tipped back to horizontal, 
\beq
Z(\pm\infty) = V_0(\pm\infty) \,.
\label{Zmfix}
\eeq
Thus we get  \emph{\DM for all coordinates}, as shown in FIG.\ref{Zmplot}~\footnote{
These formulae can actually be generalized to any constant $\fm/M$ ratio \cite{ZEBH-PR}.}.
%
%%%%%%%%%%%%%%%% FIG
\begin{figure}[h]
\includegraphics[scale=.35]{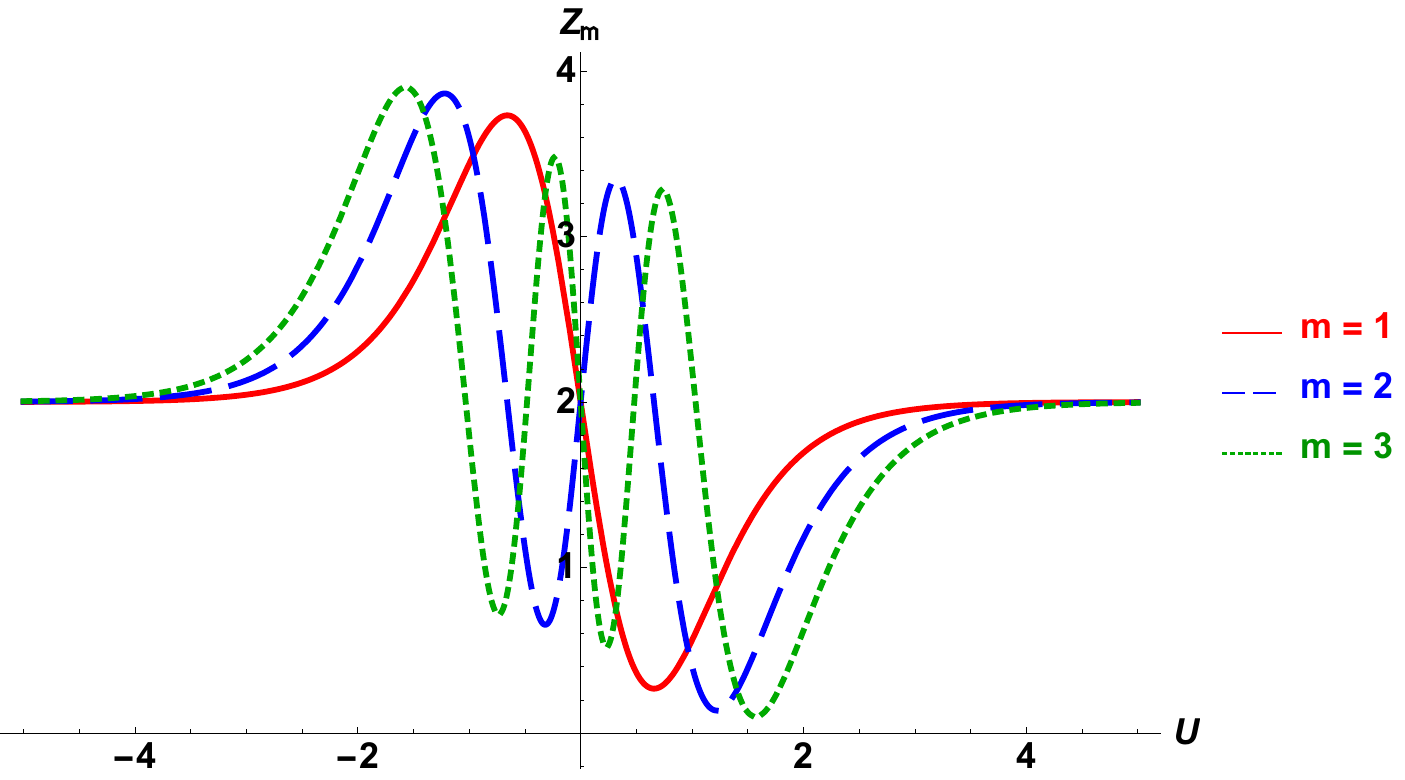}
\vskip-4mm \caption{\textit{\small The redefined coordinate $Z$ in \eqref{ZVU} exhibits DM.}
\label{Zmplot} }
 \end{figure}
%%%%%%%%%%%%%%%%%

%%%%%%%%%%%%%%%%%%%%%%%%%%%%%%%%%%
\section{Non-relativistic energy}\label{EnergySec}
%%%%%%%%%%%%%%%%%%%%%%%%%%%%%%%%%%

We conclude by studying the (transverse) energy  balance \cite{ZEBH-PR,Carneiro}. For DM parameters total  change is \emph{zero}, as seen in FIG.\ref{Ebalance}a. Off the critical amplitude we have VM, and the with vanishing outgoing velocity, FIG.\ref{Ebalance}b, implying increased energy. See also in  FIG.\ref{allkEner}.

%%%%%%%%%%%% Fig Ebalance
\begin{figure}[h]
\includegraphics[scale=.33]{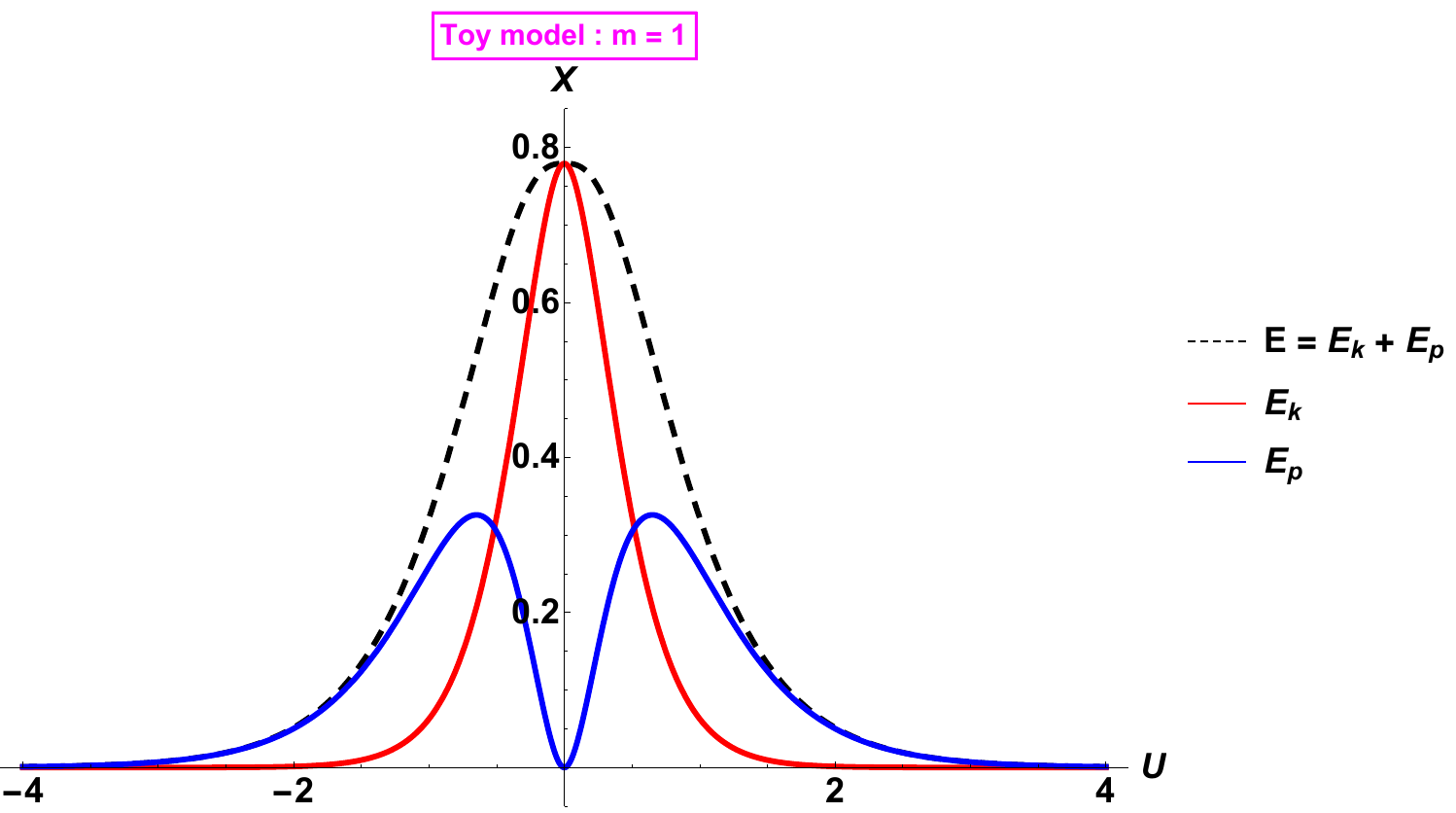}\hskip-6mm
\includegraphics[scale=.33]{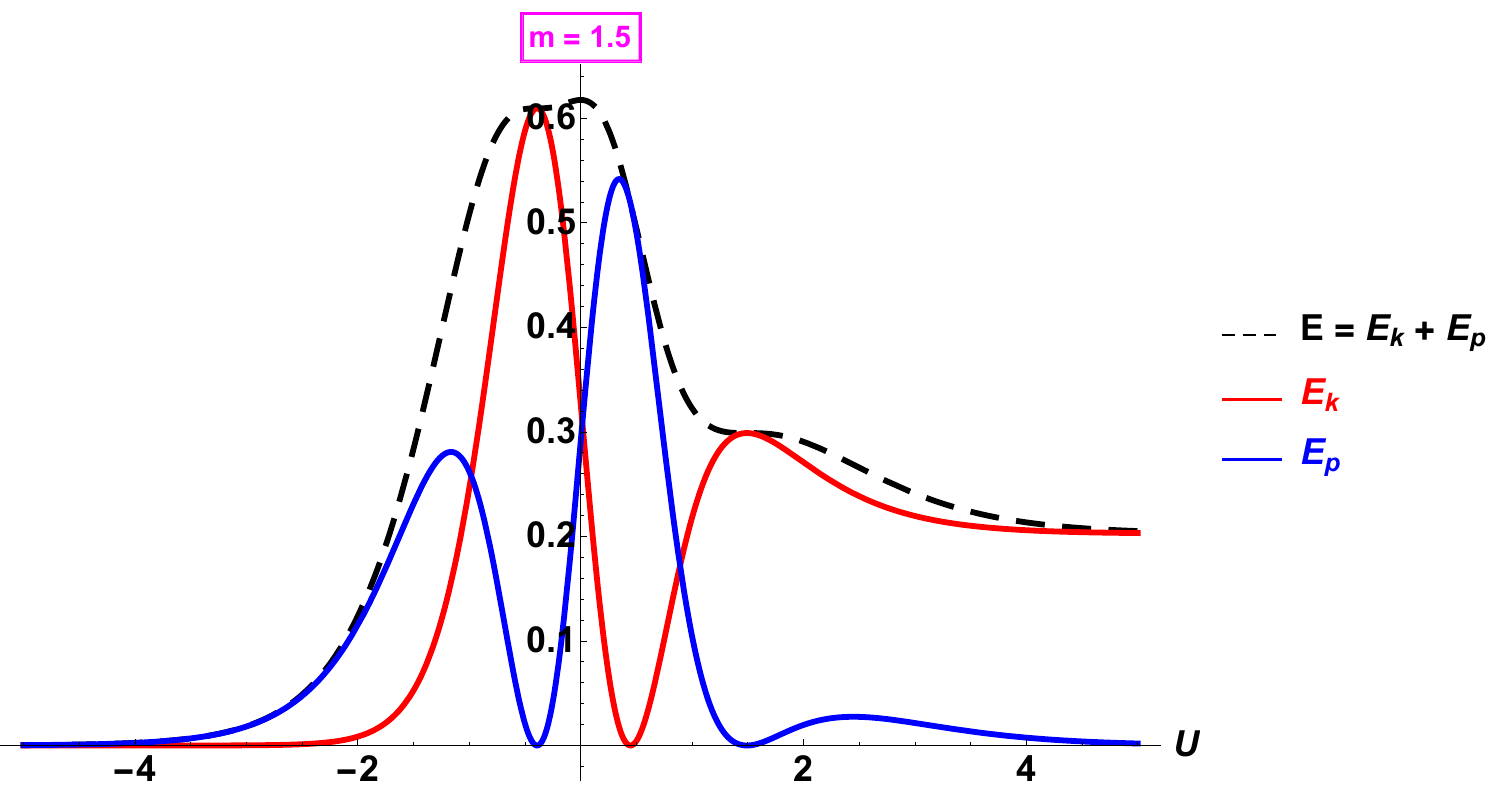}

\vskip-3mm
\hskip-18mm(a)\hskip72mm (b)
\vskip-4mm
\caption{\textit{\small For the Gaussian toy model \eqref{extoy}  the particle has zero total energy balance for DM, and positive balance for VM. }
\label{Ebalance}
}
\end{figure}
%%%%%%%%%%%
consistently with FIGs. \#15 and \#6 in \cite{DM-2}. 
\begin{figure}[h]
\includegraphics[scale=.33]{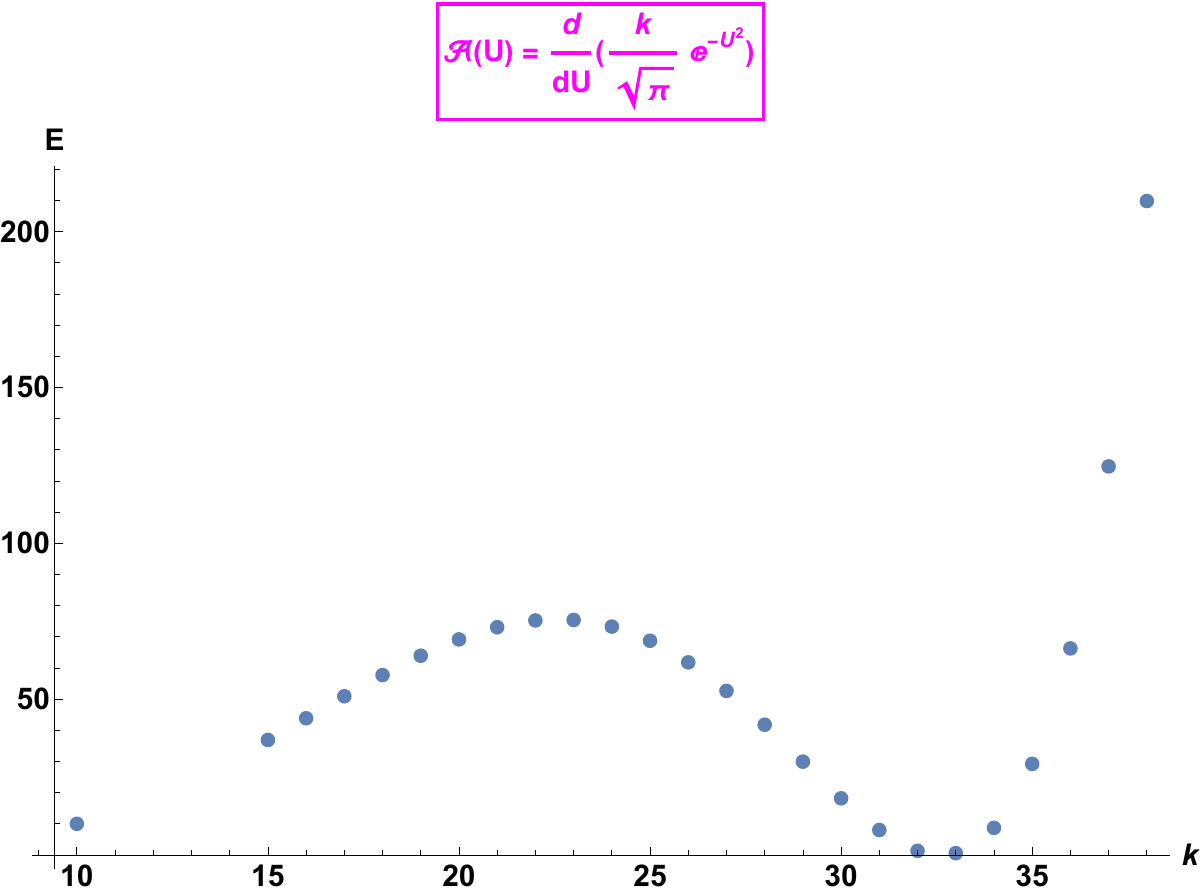}
\qquad
\includegraphics[scale=.33]{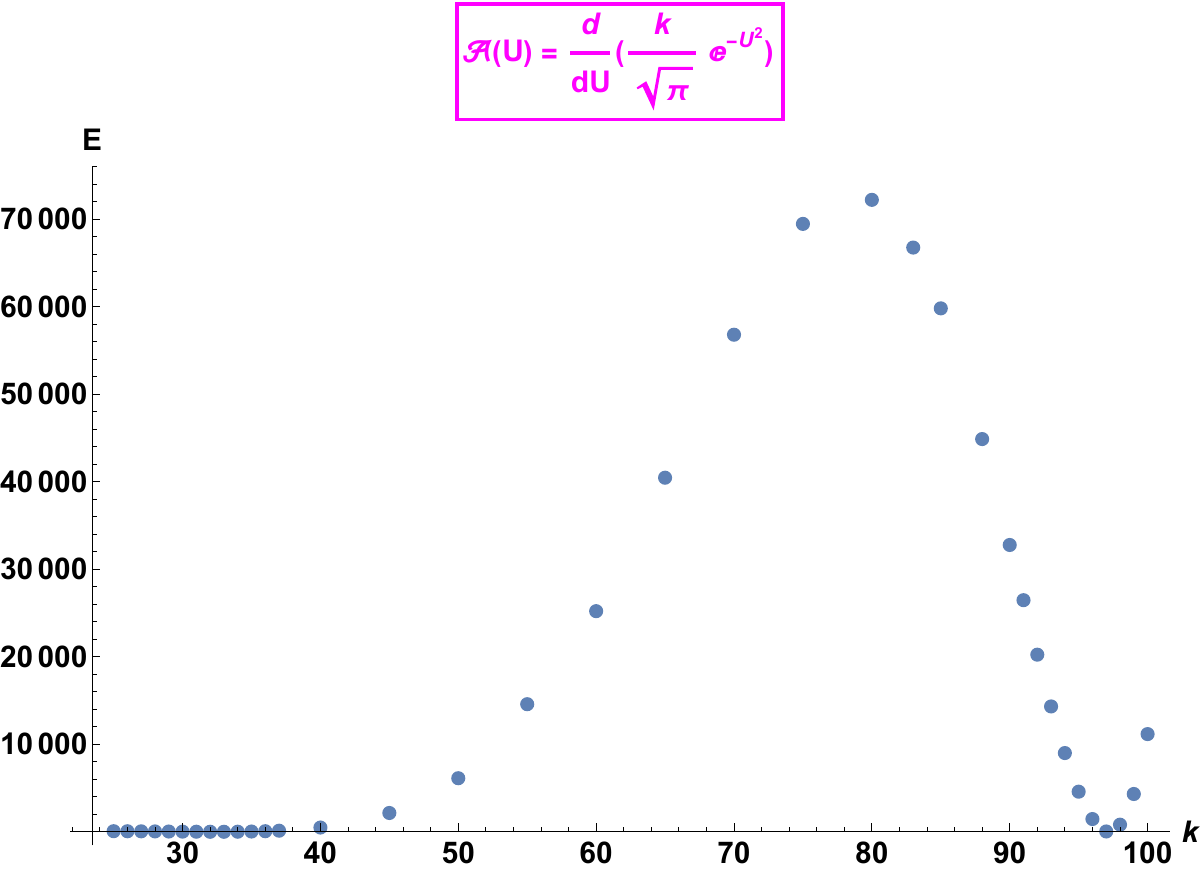}
\\
\caption{\textit{\small Off the critical amplitude $k\neq k_{crit}$, the outgoing velocity does not vanish, implying increased transverse particle energy. 
}
\label{allkEner}
}
\end{figure}

Analytic formulas were found for full \PT \cite{DM-1,DM-2},
{\small 
%%%%%%%
\besub
\begin{align}
&\hskip-6mm X_{m=1}(U)= \tanh (U)  \;
&E_{m=1}
= \frac{1}{2} \cosh(2U) \,{\rm sech}(U)^4
\label{m1PTenergy}
\\
&\hskip-6mm X_{m=2}(U)= \frac{1}{2}\Big(3\tanh(U)^2-1\Big)
& \;E_{m=2}
= \frac{3}{8}\Big(3-2\cosh(2U) + \cosh(4U)\Big)\,{\rm sech}(U)^6
\label{m2PTenergy}
\end{align}
\esub
%%%%%%%%
}

%%%%%%%%%%%%%%%%%%%%%%%%%%%%%%%%%%%%%   
\section{Sturm-Liouville equation}\label{StLSec}
%%%%%%%%%%%%%%%%%%%%%%%%%%%%%%%%%%%%%

The  $D=1$ dimensional \StL eqn \eqref{geoX2},
\beq
\dfrac {d^2\! X}{dU^2} + \cA(U) X = 0\,,
\label{geoXBIS}
\eeq 
has two independent solutions. 
The first one we denote by $P$ is required to satisfy the initial conditions : 
\beq
P(-\infty)= 1
\aand
\frac{dP}{dU}(-\infty) = 0\,,
\label{Pinitcond}
\eeq
cf. \eqref{Xinitcond}. A 2nd, independent solution which does not satisfy \eqref{Pinitcond} can be found  in terms of the \emph{Souriau matrix} \cite{Sou73,Carroll4GW,GlobalCarroll},
\beq
 Q(U) =  P(U) S(U)\,
\where 
S(U) = \displaystyle\int^{U}_{U_0}\!\!{dv}{P^{-2}(v)}\,,
\label{Smatrix}
\eeq
  see FIG.\ref{PT-PQS}. Similar figures were proposed for the approximate model \cite{Approxi}. For $D\geq2$,
  $Q$, $P$, $S$ are matrices, and  \eqref{Smatrix} generalises the scalar expression considered by Arnold in the isotropic case
 \cite{Arnold}.
 %%%%%%%%%%%%% FIG PT-PQS
\begin{figure}[h]
\includegraphics[scale=.33]{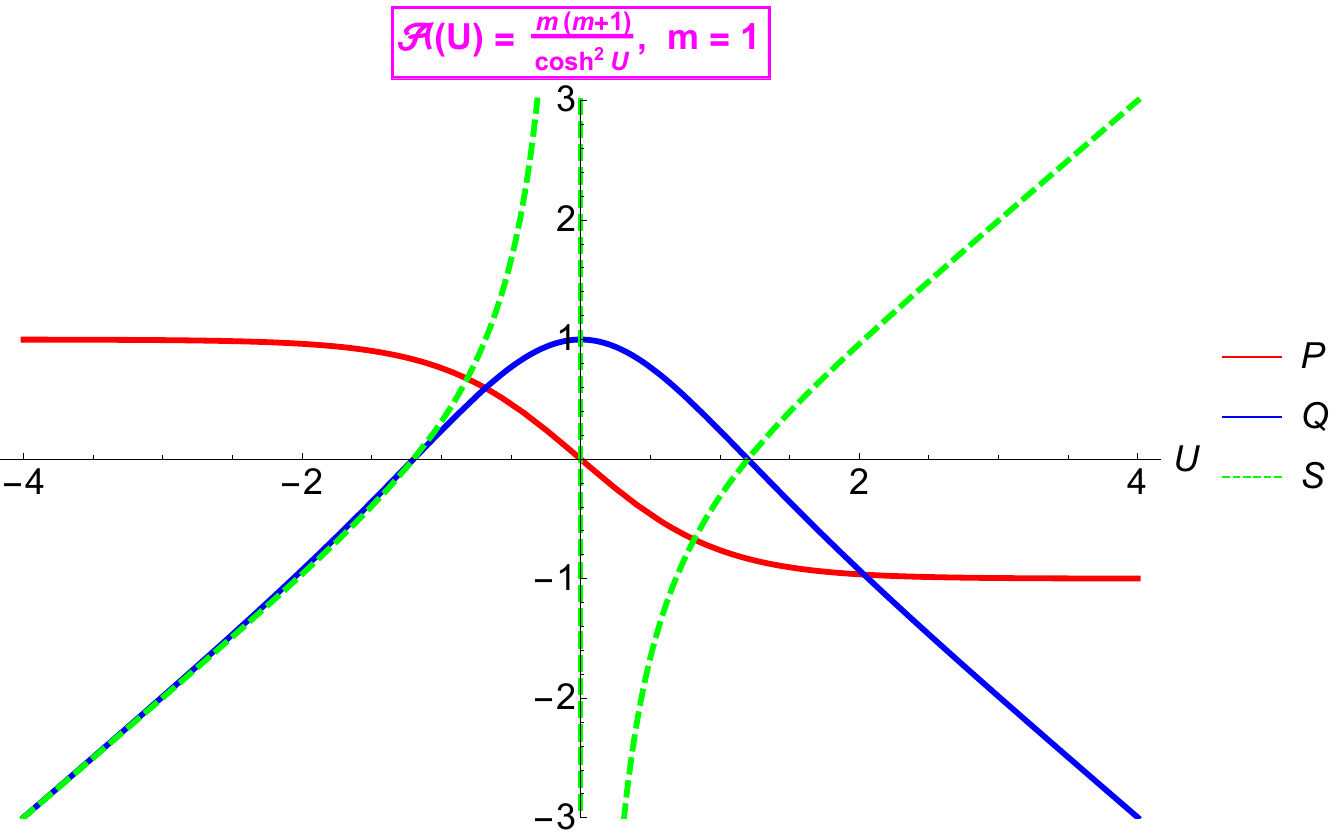}\quad\;
\includegraphics[scale=.33]{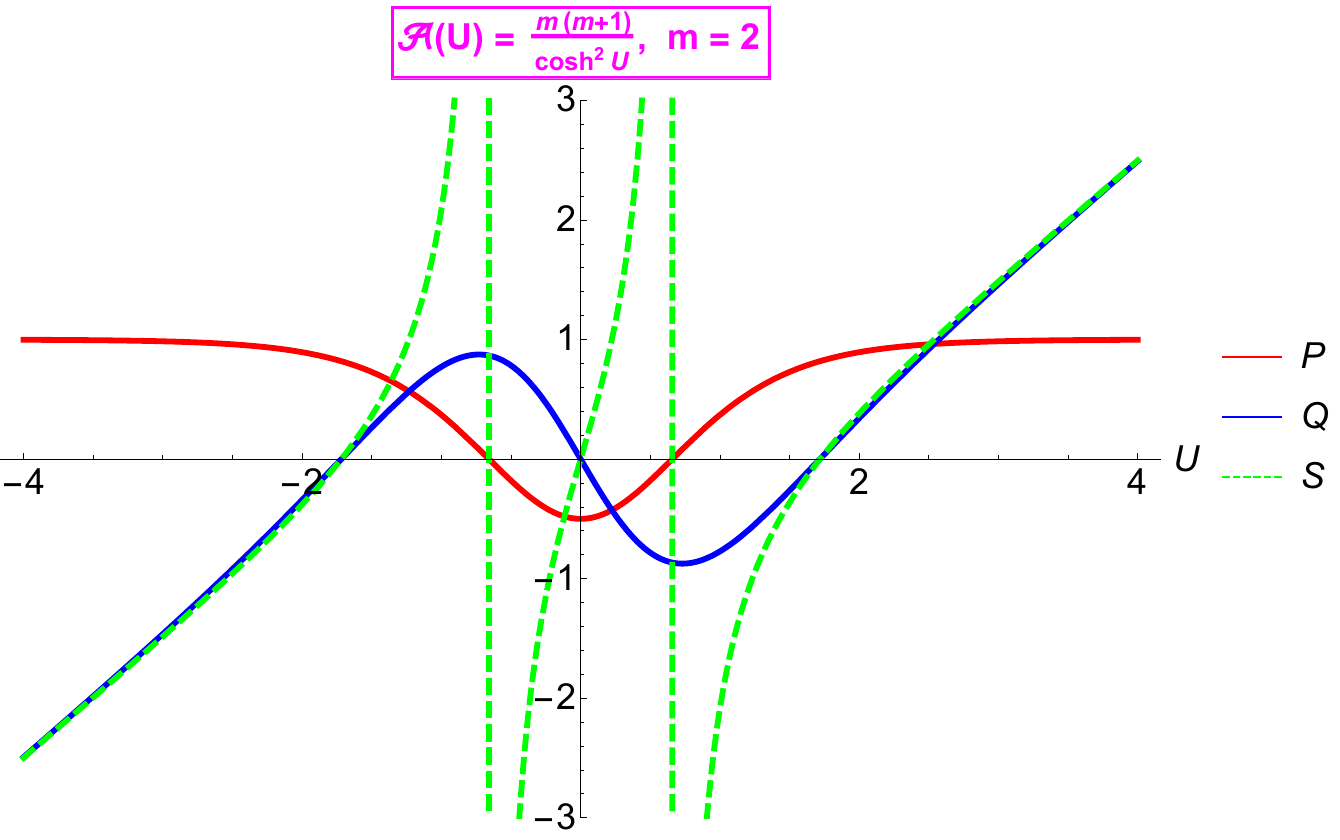}

%\\
%\vskip-2mm\hskip-11mm(a)\hskip76.5mm (b)
\vskip-3mm\caption{\textit{\small
The \StL solutions \red{${\bf P}$},
the non-DM 2nd solution \green{${\bf Q=PS}$}
and the \blue{{\bf Souriau matrix} ${\bf S}$},
 shown here for the \PT profile
  with  DM amplitude $k=k_{crit}$.
}
\label{PT-PQS} 
}
\end{figure}
%%%%%%%%%%%%%%%%%%

%%%%%%%%%%%%%%%%%%%%%%%%%%%%%%%%%%%%%%%%%%
\section{Carroll symmetry}\label{CarrolSec}
%%%%%%%%%%%%%%%%%%%%%%%%%%%%%%%%%%%%%%%%%%

 The two independent solutions
$P$ and $Q$ above  span a \emph{Carroll symmetry} \cite{LeblondCar,Sou73,Carroll4GW,GlobalCarroll}, generated by
%%%%%%%%%%%%%%%%%
\begin{equation}
h \frac{\partial}{\partial V} +
\underbrace{c \left(P \frac{\partial}{\partial X} - P' X \frac{\partial}{\partial V}\right)}_{translations}
+ \underbrace{b \left(Q  \frac{\partial}{\partial X} - Q' X  \frac{\partial}{\partial V}\right)}_{Carroll\, boosts}\,.
\label{CarrollBrink}
\end{equation}
%%%%%%%%%%%%%%%%%%
 The associated conserved quantities $\bp_0$ and $\bk_0$
 determine all geodesics \cite{GlobalCarroll}.
%%%%%%%%%%%%%%%%%%%%%%%%
\beq\medbox{
\bX(U) = P(U)\, \bk_0 + Q(U) \,\bp_0 \,.}
\label{XPQ}
\eeq
%%%%%%%%%%%%%%%%%%%%%%%%

Our approximation scheme is not limited to \PT. One can consider, for example, a Gaussian profile  \cite{Chakra,DM-1}
%%%%%%%%%%%%%
\begin{equation}
\cA^{Gauss}(U) \; \propto\,e^{-U^{2}}\,,
\label{Gaussprof}
\end{equation}
%%%%%%%%%%%%%%%%
whose shape can be made similar to \PT\! by fine-tuning, as seen in  FIG.\#6 of ref. \cite{DM-1}.
 For large $|U|$, the profile can again be approximated by a  toy profile, see FIG.\ref{ToyGaussprof}. Gaussian and \PT geodesics  are similar  \cite{DM-1,Approxi},
despite their different critical amplitudes.

%%%%%%%%%%%%%%%%%%% FIG ToyGauss
\begin{figure}[h]
\includegraphics[scale=.88]{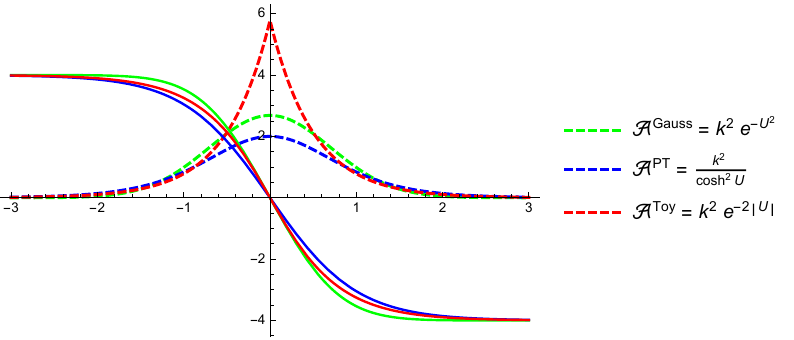}

\vskip-5mm\caption{\textit{\small
For suitably chosen parameters, the Gaussian profile and trajectory can  be approximated by that of \PT, as shown for $m=1$.
}
\label{ToyGaussprof} }
\end{figure}
%%%%%%%%%%%%%%%%%%%
 
Our approximate scheme works also for square profiles, \cite{Kar3}, shown in FIG.\ref{GaussToy}.
\goodbreak

%%%%%%%%%%%%%%%%%%% FIG GaussToy
\begin{figure}[h] 
\includegraphics[scale=.32]{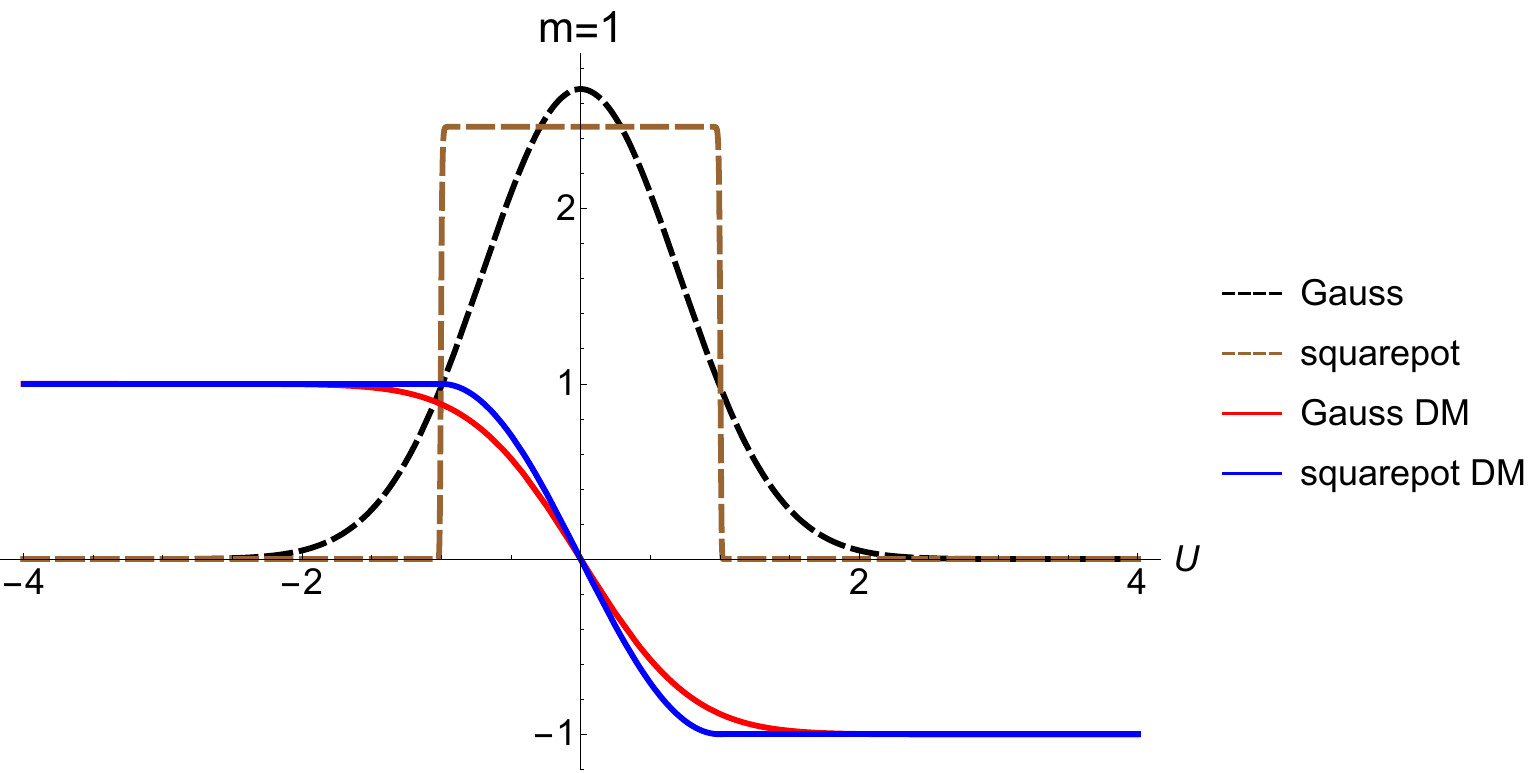}\hskip-5mm
\includegraphics[scale=.32]{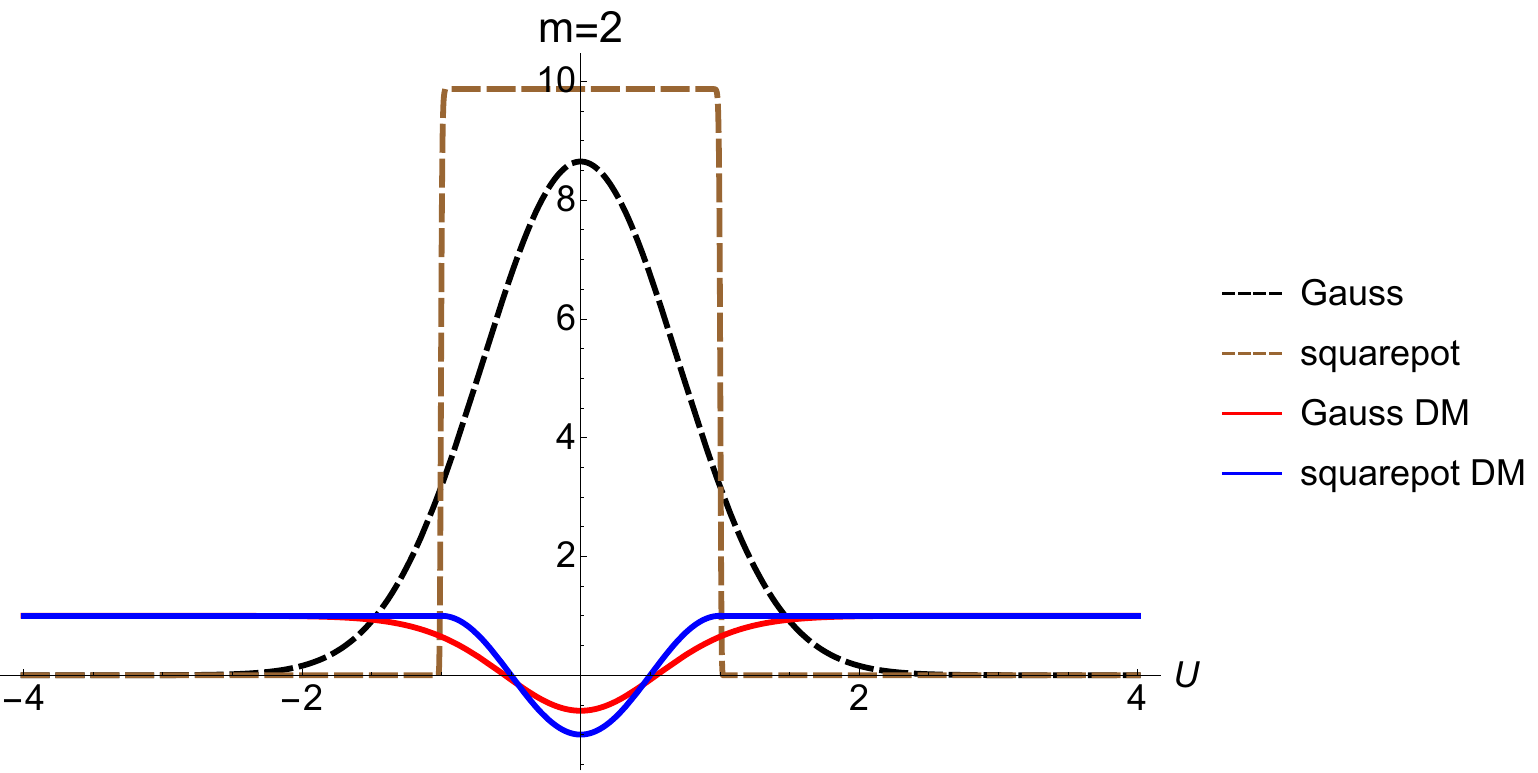}

%\\
%\vskip-2mm\hskip-20mm(a)\hskip75mm (b)
\vskip-5mm
\caption{\textit{\small  Gaussian \eqref{Gaussprof} and square profiles yield slightly differing trajectories.}
\label{GaussToy} }
\end{figure}
%%%%%%%%%%%%%%%%%%%
\goodbreak

%%%%%%%%%%%
Square profiles could also be combined antisymmetrically \cite{Benin}, yielding a double-square approximation for the flyby profile \cite{ZelPol,DM-2},
%%%%%%%%%%%%%
\begin{equation}
\cA \equiv \cA^{G}=
\frac{\;\;\,d}{dU}\left(\frac{k}{\sqrt{\pi}}%
e^{-U^{2}}\right)\,.
\label{flybyprof}
\end{equation}%
%%%%%%%%%%%%%%%%%%
 The DM trajectories are depicted in FIG.\ref{D2dsquareTraj}.
 
%%%%%%%%%%%%%%% FIG doublesquare
\begin{figure}[h]
\includegraphics[scale=.22]{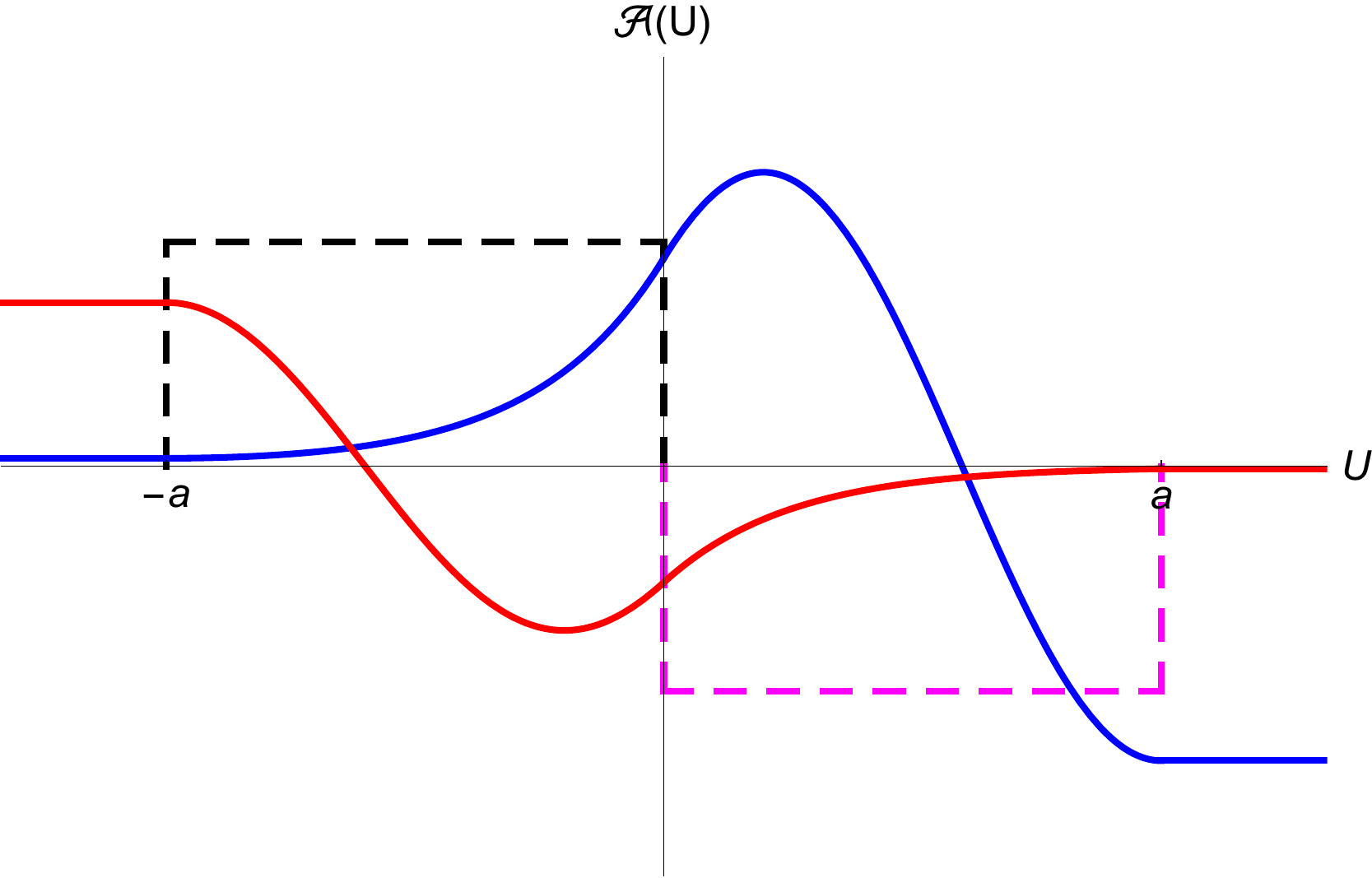} \qquad\quad
\includegraphics[scale=.22]{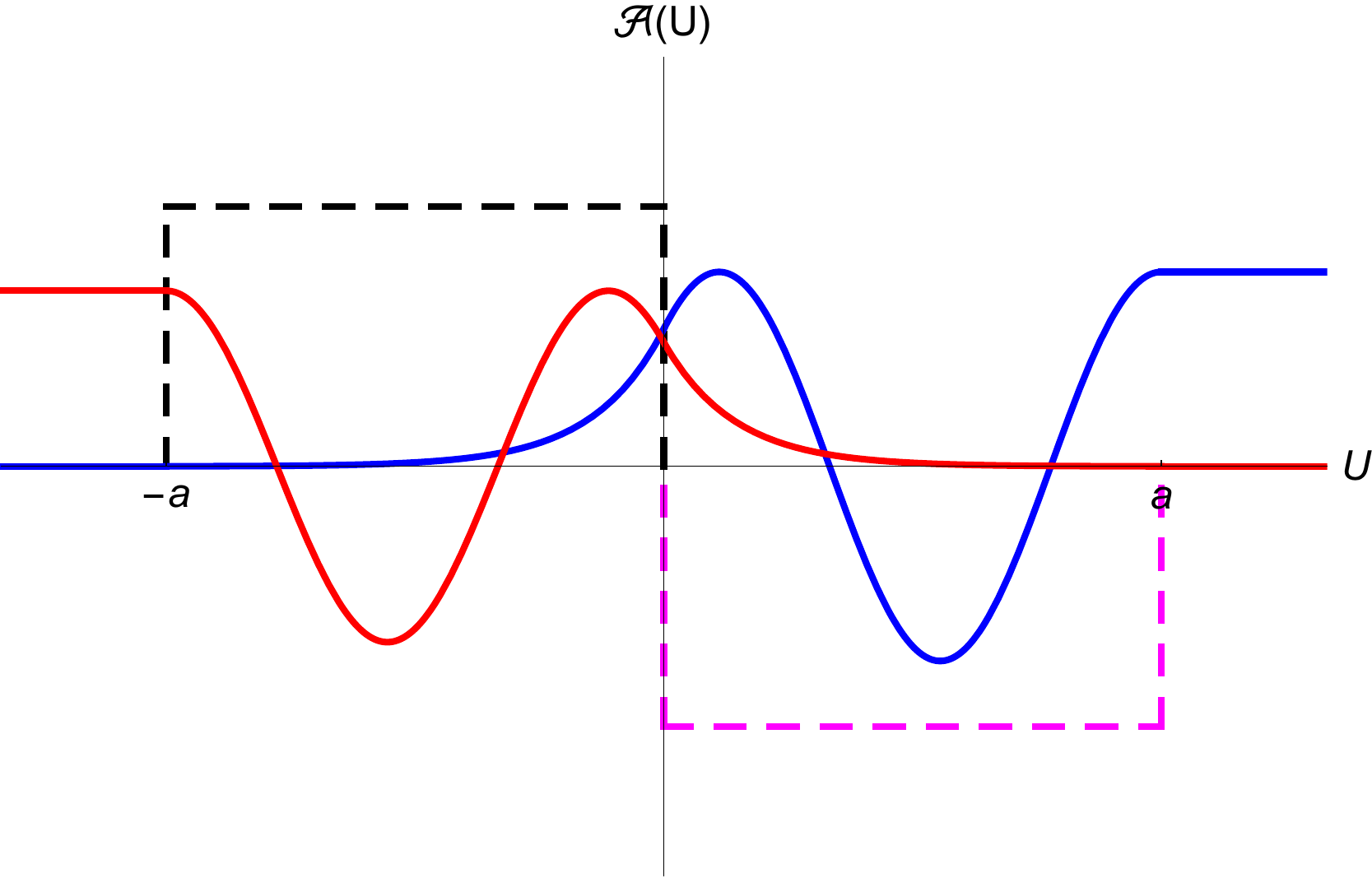}
%\vskip-3mm
%\hskip-2mm(a)\hskip67mm(b)

\vskip-6mm
\caption{\textit{\red{$D=2$} double-square approximation of the flyby profile \eqref{flybyprof} with wave numbers ${\bf m=1}$ and ${\bf m=2}$.} 
\label{D2dsquareTraj}
}
\end{figure}
%%%%%%%%%%%%

%%%%%%%%%%%%%%%%%%%%%%%%%%%%%%%%
\section{Limitations of the sandwich approximation}\label{LimitMemSec}
%%%%%%%%%%%%%%%%%%%%%%%%%%%%%%%

From a dynamical standpoint, the sandwich condition requires the test particle to move freely both before the arrival and after the passage of the gravitational wave. This implies that the profile $\mathcal{A}(U)$ must decay sufficiently rapidly so as not to affect the scattering data -- namely, the outgoing velocity and the intercept should remain well defined, 
\begin{eqnarray}
	X(+\infty)=a+bU, \quad a<\infty, \quad b<\infty\,.
	\label{free-outgoing}
\end{eqnarray}

The ME in a sandwich wave is then described by these two scattering data:  VM is determined by the outgoing velocity $b$, while DM is characterized by the intercept $a$ when the outgoing velocity vanishes, $b=0$. This has been well illustrated in our earlier examples.

However, a subtle and easily overlooked point is that: \emph{while the sandwich condition necessarily leads to ME, the latter does not necessarily require the sandwich condition,} which is merely  sufficient, but not necessary 
 for  ME \cite{ZhaoCao}. This is because the essence of the ME lies in identifying a constant change in the dynamical parameters: for VM, this merely means that the outgoing velocity undergoes a constant shift, while the trajectory may be allowed to have a small drift whose derivative vanishes as $U\to\infty$, 
\begin{eqnarray}
	\frac{dX_{\rm VM}}{dU}(+\infty)=b<\infty \Rightarrow X_{\rm VM}(+\infty)=bU+f(U) \quad \mathrm{where} \quad \frac{df}{dU}(+\infty)=0\,.
	\label{VM-outgoing}
\end{eqnarray}
For DM, even smaller drift of the trajectory is allowed,
\begin{eqnarray}
	X_{\rm DM}(+\infty)=a+o(1).
	\label{DM-outgoing}
\end{eqnarray} 

A next natural question is: how do we find the limit of the sandwich condition? The answer lies in the asymptotic motion of the particle,  determined by the decaying tail of the profile $\mathcal{A}(U)$.
The general outgoing asymptotic motion can be written as \cite{ZhaoCao}
\begin{eqnarray}
	X(+\infty)=X_{\rm free}+\delta X+O(\mathcal{A}^{2}), \quad
	X_{\rm free}=a+bU, \quad
	\delta X=\delta a +\delta b \, U,
\end{eqnarray}
where 
\begin{eqnarray}
	\delta a=\int_{U_*}^{\infty}\mathcal{A}(s)X_{\rm free}(s)ds, \quad
	\delta b=\int_{U_*}^{\infty}s\mathcal{A}(s)X_{\rm free}(s)ds,
\end{eqnarray}
are corrections to the intercept and the outgoing velocity, respectively, controlled by the decay of the profile tail. Here $U_*$ is a large but finite value, after which the profile enters its decaying tail region.

If the asymptotic motion retains the free form \eqref{free-outgoing}, then the sandwich condition still holds, and the integral corrections to the outgoing velocity and intercept should all converge. This requires the decaying tail of the profile to satisfy the following integral condition:
\begin{eqnarray}
	\int_{U_*}^{\infty}s^2\mathcal{A}(s)ds<\infty.
\end{eqnarray}
Such a condition is also the criterion for the profile to satisfy the sandwich condition, and  yields a critical decay behavior: $\mathcal{A}(U)\sim U^{-3}$. When the decay is faster than this, the sandwich condition still holds. When the decay is equal to or slower, then the outgoing motion is no longer free and the sandwich condition is broken; however the ME may still persist. We illustrate this point by the Scarf and the inverse-cubic profiles.

The Scarf profile is given by the function \cite{Sila-PLB}
\begin{eqnarray}
	\mathcal{A}^{\rm Scarf}=k\frac{\sinh U}{\cosh^2U}\,,
\end{eqnarray}
whose tail behaves as $\mathcal{A}(U)\sim e^{-U}$, satisfying sandwich condition.

By contrast, the inverse-cubic profile is described by \cite{ZhaoCao}
\begin{eqnarray}
	\mathcal{A}^{\rm IC}=\frac{kU}{(U^2+1)^2},
\end{eqnarray}
whose decaying tail is $\mathcal{A}(U)\sim U^{-3}$. Such profile breaks the sandwich condition. Consequently, the asymptotic motion will no longer be free. However, the memory effect still exists.

FIG. \ref{Scarf-U3-VM} illustrates  VM for the Scarf  and the inverse-cubic profiles. The outgoing motion in the Scarf profile remains strictly free, whereas the outgoing motion in the inverse-cubic profile deviates from the free straight line at large $U$ \cite{ZhaoCao}, so it is no longer free. However, the outgoing velocities in both cases remain constant, so VM still persists.

\begin{figure}[h]
	\includegraphics[scale=.35]{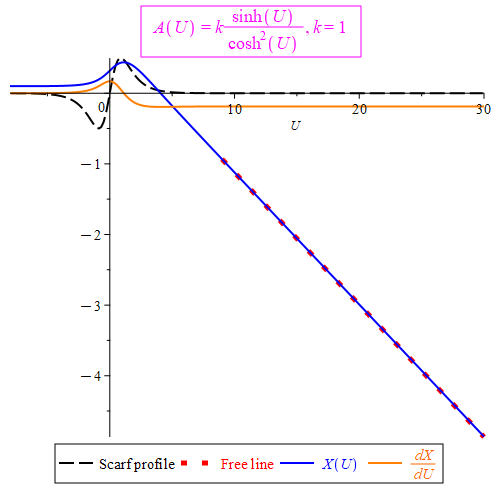} \qquad\quad
	\includegraphics[scale=.36]{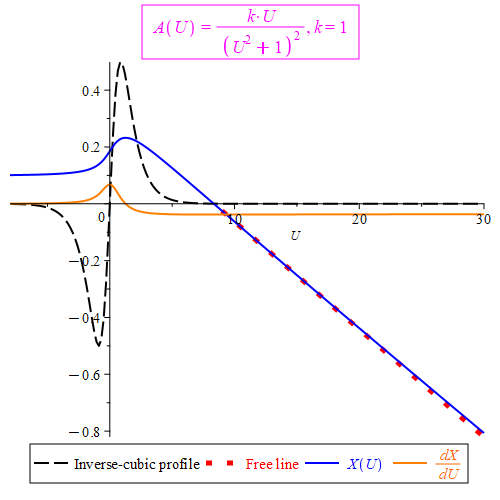}
	%\vskip-3mm
	%\hskip-2mm(a)\hskip67mm(b)
	
	\vskip-5mm
	\caption{\textit{These two plots show VM motions for the Scarf  and the inverse-cubic profiles, respectively. The \blue{blue} and \orange{orange} curves describe the outgoing motion {$X(U)$} and its velocity {$\frac{dX}{dU}$}, the \red{red} line is the {free straight line}}.
		\label{Scarf-U3-VM}
	}
\end{figure}

The displacement memory effect can still arise in both these two profiles. In the Scarf case, it appears when \cite{Sila-PLB}
\begin{eqnarray}
	k=(2n+1)\sqrt{n(n+1)}\,,
\end{eqnarray}
while for the inverse-cubic profile, it also occurs when \cite{ZhaoCao}
\begin{eqnarray}
	k=2(n+1)\sqrt{n(n+2)}\,.
\end{eqnarray}
These DM motions are shown in FIG. \ref{Scarf-U3-DM}.

\begin{figure}[h]
	\includegraphics[scale=.36]{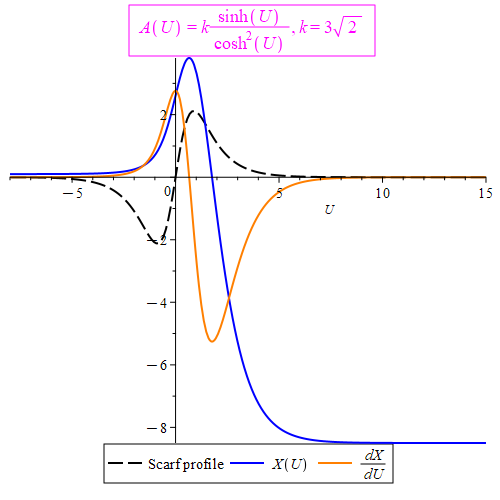} \qquad\quad
	\includegraphics[scale=.36]{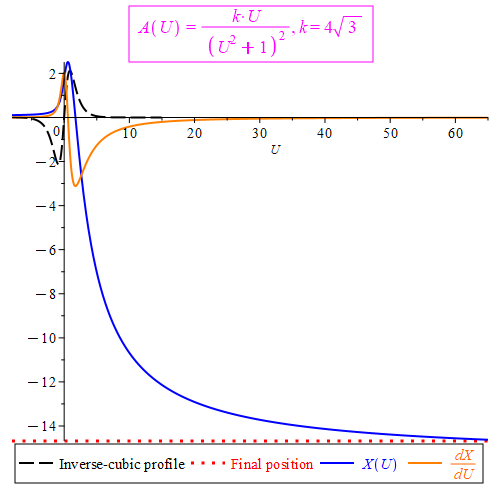}
	%\vskip-3mm
	%\hskip-2mm(a)\hskip67mm(b)
	
	\vskip-5mm
	\caption{\textit{These two plots show DM motions for the Scarf  and the inverse-cubic profiles, respectively. The \blue{blue} and \orange{orange} curves describe the outgoing motion {$X(U)$} and its velocity {$\frac{dX}{dU}$}, the dotted \red{red} horizontal line indicates the final position of the outgoing motion in the inverse-cubic profile.}
	\label{Scarf-U3-DM}
	}
\end{figure}

By studying the asymptotic motion, we can also identify another critical behavior, $\mathcal{A}(U)\sim U^{-2}$, which determines whether the ME does still exists: when the decay rate is faster, then the memory effect persists; when the decay rate is equal to or slower than it, the memory effect is lost. See  Ref. \cite{ZhaoCao} for details.

%%%%%%%%%%%%%%%%%%%%%%%%%%%%%
\section{Conclusion}\label{Concl}
%%%%%%%%%%%%%%%%%%%%%%%%%%%%%

\vspace{-3mm}
We reviewed current knowledge on the Memory Effect for approximate profiles. We just mention that
(super)symmetry aspects are studied in ref. \cite{SUSY}. Further related work is found also, for example, in \cite{Andrz20,Ilderton, APrencel-1,APrencel-2,A-Kar}.
We call attention at the pioneering work of Podolsky et al. \cite{Pod98,PodolskyVeselyCz98} on the impulsive limit. Related work includes  \cite{Reza-1,Reza-2,Mojtaba-postCarroll}.
\goodbreak

\begin{acknowledgments} 
PMZ was partially supported by the National Natural Science Foundation of China (Grant No. 12375084). PMZ,
ME, JB and PAH were supported by the Scientific and Technological Research Council of Turkey (T\"UBITAK), grant number 125F021.
\end{acknowledgments}
\goodbreak

%%%%%%%%%%%%%
%%%%%%%%%%%%%
\end{document}